\documentclass[]{IET}

\usepackage{cite}
\usepackage{algorithm}
\usepackage{algorithmic}
\renewcommand{\algorithmicrequire}{\textbf{Input:}}
\renewcommand{\algorithmicensure}{\textbf{Output:}}

\usepackage{graphicx}
\usepackage{upgreek}
\usepackage{graphics}

\usepackage{float}
\usepackage{subfigure}
\usepackage{array}
\usepackage{multirow} 
\usepackage{amsmath}
\usepackage{xcolor}
\usepackage{booktabs}
\usepackage{url}
\usepackage{lettrine}

\begin{document}

\title{Anomaly Detection and Redundancy Elimination of Big Sensor Data in Internet of Things}

\author{Sai Xie}

\author{Zhe Chen}
\affil{School of Computer Science and Engineering, Northeastern University, China}

\affil{Email: chenzhe@mail.neu.edu.cn}

\abstract{In the era of big data and Internet of things, massive sensor data are gathered with Internet of things. Quantity of data captured by sensor networks are considered to contain highly useful and valuable information. However, for a variety of reasons, received sensor data often appear abnormal. Therefore, effective anomaly detection methods are required to guarantee the quality of data collected by those sensor nodes. Since sensor data are usually correlated in time and space, not all the gathered data are valuable for further data processing and analysis. Preprocessing is necessary for eliminating the redundancy in gathered massive sensor data. In this paper, the proposed work defines a sensor data preprocessing framework. It is mainly composed of two parts, i.e., sensor data anomaly detection and sensor data redundancy elimination. In the first part, methods based on principal statistic analysis and Bayesian network is proposed for sensor data anomaly detection. Then, approaches based on static Bayesian network (SBN) and dynamic Bayesian networks (DBNs) are proposed for sensor data redundancy elimination. Static sensor data redundancy detection algorithm (SSDRDA) for eliminating redundant data in static datasets and real-time sensor data redundancy detection algorithm (RSDRDA) for eliminating redundant sensor data in real-time are proposed. The efficiency and effectiveness of the proposed methods are validated using real-world gathered sensor datasets.}

\maketitle

\section{Introduction}
Nowadays, Internet of things (IoT) has gradually integrated into our lives. The challenge of deriving insights from IoT has been recognized as one of the most important opportunities for both academia and industry. The basic idea of IoT is to connect all things by the Internet. It is expected that things can be identified automatically, can communicate with each other, and even can make decisions by themselves\cite{Tsai2014Data}. The development of computer technology makes lots of IoT application come into reality. IoT and machine-to-machine were worth $\$44.0$ billon in 2011 and are expected to grow up to $\$290.0$ billion by 2017\cite{Tsai2014Future}.

In IoT systems, many sensors are embedded into equipment and machines. These sensors can collect different types of sensor data, such as environmental data, traffic data, and logistic data. So the data gathered by IoT have the following features\cite{Chen2014Big}:
\begin{itemize}
    \item Large-scale: Massive sensor data are gathered by distributed equipments. There are plenty of sensor data generated everyday. In order to analyze and process the data, all of these data should be stored within a certain period. Therefore, the data generated by IoT is large-scale.
    \item Heterogeneity: In the system of IoT, there are variety of data acquisition devices. The type of gathered data is also different. The devices are heterogenous, too. Thus, all of these factors cause IoT data to be heterogenous.
    \item Strong time and space correlation: Sensor data of IoT gathered by devices that are placed at specific locations are labeled with time stamps. And data streams are measurements of continuous physical phenomenon. Spatial and temporal correlations within data streams are inherent. Thus, time-space correlation is one of the most critical property of data gathered by IoT.
\end{itemize}

As mentioned above, IoT data is actually one type of big data. There are heterogeneous data sources and data types to represent the data. Data generated from IoT are considered to contain highly useful and valuable information.
\subsection{Motivation}
    \begin{figure}[!t]
	\centering
	\includegraphics[height=3cm,width=8cm]{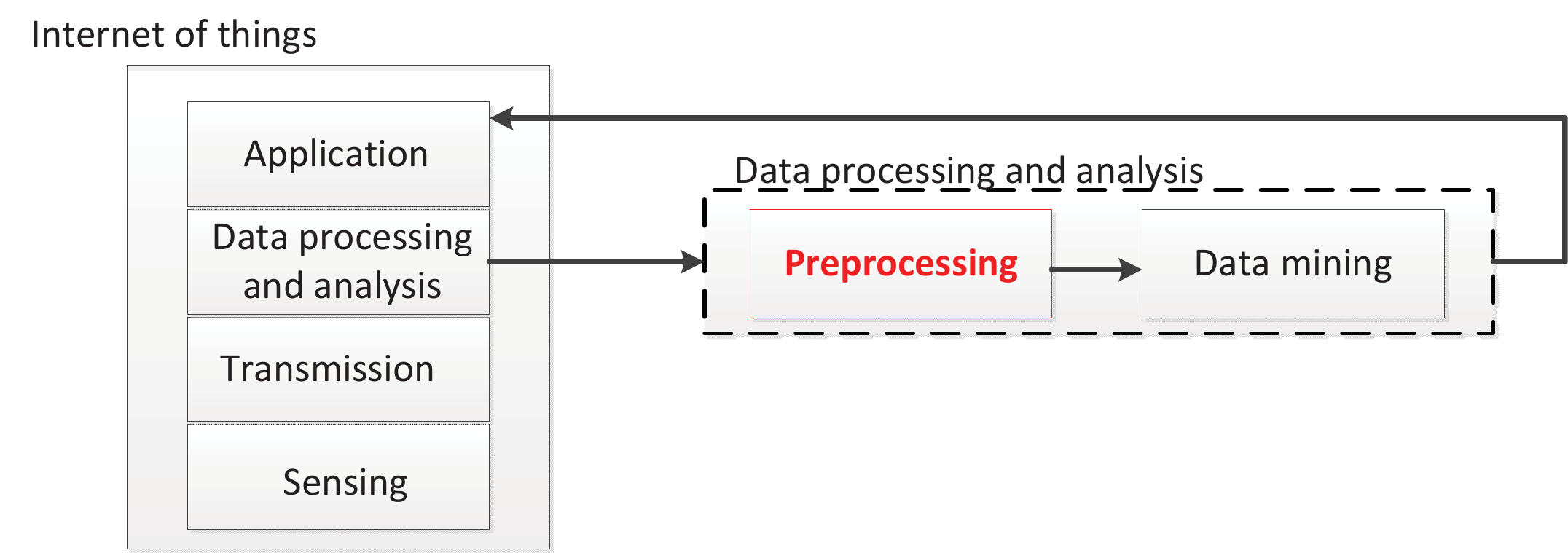}
	\caption{The architecture of IoT with data mining}
	\label{fig:preprocessing}
    \end{figure}
As big sensor data collected by IoT, how to handle these data and how to dig out useful information play an important role in IoT systems. Data analysis for sensors and devices not only helps us grasp running status, but also helps us make optimal decisions. But various reasons, such as data anomalies, redundancy, and data missing may fail the collected data to be directly used. Thus, it is necessary for sensor data to be preprocessed. As shown in Fig. \ref{fig:preprocessing}, the structure of Internet of things mainly consists of four layers. And for data processing and analysis layer there may consist of two sublayers. It can be seen that the data preprocessing sublayer is responsible for the big sensor data gathered by IoT and it feeds the extracted valuable data to data mining sublayer. It can be seen that data preprocessing is one of the most critical step in the process of data mining.

Because of the characteristics of sensor nodes in IoT system, the probability that the data sampled by a node is highly correlated or repetitious over time is quite high \cite{fateh2013energy}. So it is normal that the collected data-sets contains redundancies. And anomaly data is also one of the most common phenomenon that appears in the collected data-sets. Currently, for sensor data preprocessing in IoT there is less reported research on anomaly detection and redundancy elimination. And these two parts are also indispensable for sensor data preprocessing. Therefore, the main purpose of this paper focuses on anomaly detection and redundancy elimination of big sensor data.
\subsection{Related Work}
\subsubsection{Sensor Data Anomaly Detection}
Anomaly detection is the problem of finding patterns in data that do not conform to an a priori expected behavior \cite{marti2015anomaly}. Paper \cite{illiano2015detecting} reviews the related work and derives general principles and a classification of approaches within this domain. Based on graph theory and exploiting spatiotemporal correlations of physical processes, a fully distributed general anomaly detection (GAD) scheme is introduced in \cite{chen2015distributed}. There are also some methods based on support vector machine (SVM) for anomaly detection as shown in \cite{Salem2014Anomaly}\cite{martins2015support}\cite{Xiao2008Attribute}. However, as we know, SVM-based anomaly detection algorithms are sensitive to missing data. In paper \cite{Shaikh2014Efficient}, the problem of distance-based outlier detection on uncertain datasets of Gaussian distribution is discussed, and a cell-based approach is proposed in this paper to quickly identify the outliers. But the choice of parameters in the algorithm has a great impact on the experimental results, so the algorithm is unstable in anomaly detection. In paper \cite{Zhao2014Adaptive}, an adaptive fuzzy clustering based anomaly data detection is proposed. paper \cite{shaikh2014top} studies the problem of top-k distance-based outlier detection from uncertain data objects. The complexity of the distance-based anomaly detection algorithm is relatively low, but the accuracy of the anomaly detection can not be guaranteed for real-world datasets. In paper \cite{Ma2016Supervised}, a novel framework that supports anomaly detection in uncertain data streams is proposed and the proposed framework adopts a wavelet based soft-thresholding method to remove noises or errors in data streams.

From the perspective of technology, traditional anomaly detection methods can be roughly divided into: distance-based methods \cite{Angiulli2009DOLPHIN} \cite{Moshtaghi2011Clustering}, density-based methods \cite{Huang2014Physics} \cite{Liu2012Isolation}, model-based methods \cite{Sch2001Estimating}, and so on. The distance-based methods mainly use some common distance formulas (such as Euclidean distance) as a measure to find abnormal data. Firstly, the distance between target sample and the center of the detection model is calculated. If the distance is greater than a preset threshold, then it is considered to be abnormal. The density-based approach is an extension of the distance-based approach. If the density of the area where target sample is located is less than the set threshold, then the sample is anomalous. Generally speaking, in local anomaly detection, the density-based approach is more accurate than other methods. In model-based approach, anomaly detection model can be learned through historical data. Normally, some statistical models or machine learning methods such as Gaussian distribution, artificial neural network, or support vector machine can be used to establish the detection model. All in all, the methods described above may have a certain overlap. And it can be seen clearly that anomaly detection based on the machine learning takes the leading position.
\subsubsection{Sensor Data Redundancy Elimination}
In the fields of Internet or IoT, redundant data can cause the problems of deteriorating the information transmission and increasing energy consumption. With the rapid growth of IoT, redundancy elimination has attracted much attention in recent years from both academia and industry.

In paper \cite{Liang2005Redundancy}, a method which uses Singular-Value-QR Decomposition (SVDQR) to reduce redundancy in wireless sensor networks is proposed. This algorithm is just to select principal data sets from particular sensor nodes to represent all the sensor nodes in the neighborhood, so the accuracy of redundant node detection is not very high. In wireless sensor network, data aggregation is usually used to eliminate redundant transmission by aggregating data from multiple sensors. In \cite{vinas2015redundancy}-\cite{Coudert2015Robust}, the solutions based on data aggregation for redundancy elimination in WSN are proposed. The main and most important improvement in these proposed solutions is based on the concept of selecting the cluster head and determining which node sends the information when redundant data are detected. In \cite{Patil2010SVM}, an SVM based data redundancy elimination for data aggregation in WSN has been proposed. Firstly, an aggregation tree for a given size of sensor network is built. Then, SVM based method is applied to the tree to eliminate redundant data. By exploiting the range of spatial correlations of data in the network, redundancy elimination for accurate data aggregation (READA) applies a grouping and compression mechanism to remove redundant data which is introduced in \cite{Khedo2010READA}.

It can be seen that most of the researches on redundant data elimination focus on data aggregation and data compression at routing or protocol level. However, there are few algorithms to eliminate redundancy directly from the perspective of data. As we mentioned above, in the era of big data and IoT, sensor data processing and analysis is one of the most critical step in the architecture of IoT. So redundancy elimination as a substep of sensor data processing and analysis in IoT is obviously one of the most important issue to be solved.
\subsection{Contribution of This Paper}
In this paper, a framework for sensor data preprocessing is proposed. The framework is composed of two parts, one for sensor data anomaly detection and the other for redundancy elimination. The contribution of this paper is trifold.

Firstly, based on the characteristics of sensor network in IoT, Bayesian network is proposed to model the problem of preprocessing of sensor data gathered by IoT. Secondly, an algorithm based on principal statistic analysis and Bayesian network is proposed for sensor data stream anomaly detection. And the features of gathered sensor data can be extracted according to the principal statistic model, then the anomaly detection of collected data stream can be conducted by the combination of extracted data feature and Bayesian network of sensor nodes. Thirdly, considering that a dynamic Bayesian networks (DBNs) is an extension of static Bayesian network (SBN) to temporal domain, condition dependencies are modeled between random variables both within and across time slots. Thus, the model of DBN can be designed for the analysis of temporal sequences. And two sensor data redundancy elimination approaches based on SBN and DBNs are proposed, respectively, i.e., static sensor data redundancy detection algorithm (SSDRDA) for eliminating redundant data in static data sets, and real-time sensor data redundancy detection algorithm (RSDRDA) for eliminating redundant sensor data in real-time.

This paper is organized as follows. The problem modeling using Bayesian networks is described in Section 2. Then, method for learning the structure of Bayesian network from the gathered sensor data is presented in Section 3. In Section 4, an algorithm for anomaly detection in big sensor data is proposed. And in Section 5, the algorithms for static sensor data redundancy detection and real-time sensor data redundancy detection are proposed. Finally, based on the gathered real-world sensor datasets, the performance analysis and evaluation of our methods are discussed in Section 6.
\section{Problem Modeling}
Big data in IoT are virtually collected by hundreds of thousands of sensor nodes. Because of the information communication of each node, a certain dependence exists among these sensor nodes. Therefore, we intend to use Bayesian network to describe the relationship among those sensor nodes.

For a specific time if the data gathered by each sensor node is regarded as a variable then we can use SBN to represent a set of variables in form of nodes on a directed acyclic graph. It indicates the conditional dependencies of the random variables. If the random variables are defined
as a sequence $X=\{x_{1},x_{2},\cdots x_{n}\}$ in SBN and $x_{i}$ is conditional dependent of its non-descendants given its parents. Therefore the joint distribution of random variable $x_{i}$ can be
written as $P(x_{1},x_{2},\cdots,x_{n})=\prod_{i=1}^{n}P(x_{i}|pa(x_{i}))$, where $pa(x_{i})$ is the parent of $x_{i}$.

\begin{figure}[!t]
\centering 
\includegraphics[height=3cm,width=8cm,scale=0.5]{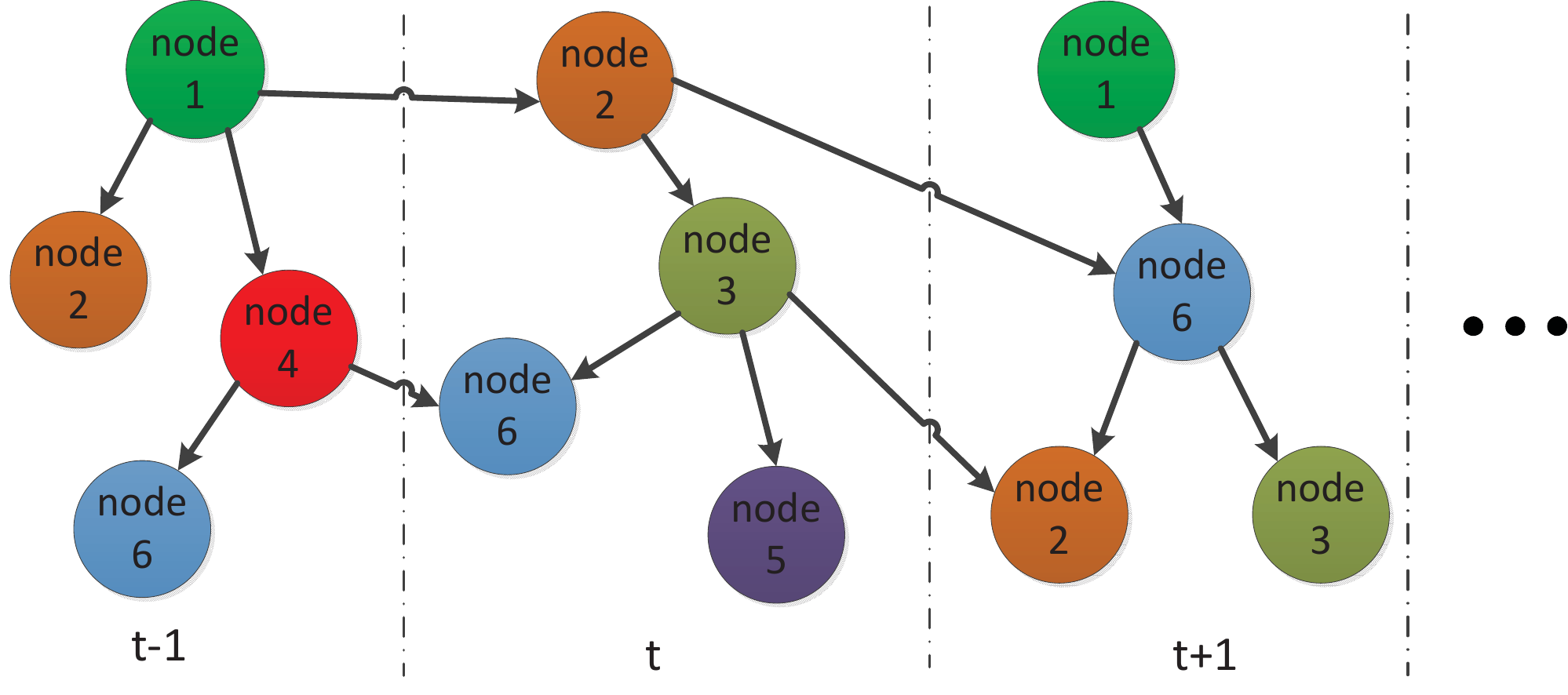}
\caption{A DBN model for sensor nodes at different time (different color denotes different state of sensor nodes)} 
\label{fig:dbn}
\end{figure}

Fig. \ref{fig:dbn} shows a dynamic Bayesian network model for sensor nodes at different time. It can be seen that the working state of sensor nodes is constantly changing at different moments. And there may be some nodes exiting or joining the sensor network at any time. So the dependencies of the sensor nodes in a certain period of time and the dependencies of the nodes between two time slices are constantly changing. A DBN is an extension of SBN to time domain. Because the characteristic with time epoch in a DBN, it is suit for dealing with real-time problem. However, building a DBN with lots of random variables is a complex project. In practice, we assume that the structure of a Bayesian network for sensor nodes will not change sharply in a limited period of time. Thus, in order to simplify this problem, we make some reasonable assumptions\cite{Ghanmy2011Characterization}:
\begin{enumerate}
\item[*] The variation of condition probability is stable at a specific time.
\item[*] A dynamic process can be modeled by a first-order-Markovian. $$P(x[t+1]|x[1],x[2],\cdots,x[t])=P(x[t+1]|x[t])$$
\item[*] The transition probability $P(x[t+1]|x[t])$ is stable in a time slot $t$.
\end{enumerate}

A DBN is formed with two parts $(B_{0},B_{\rightarrow})$, where $B_{0}$ is initial network which defines the prior $P(x[0])$, $B_{\rightarrow}$ is a transition network that defines a two slice temporal Bayes net\cite{Sun2015A}. And it can be learned the relationship of nodes at current temporal from initial network. The relationship of nodes between two slice temporal can be obtained from transition network. The joint distribution of the model of DBN in Fig.\ref{fig:dbn} can be gotten by unrolling two slice temporal Bayes net till the network has $T$ slice and by multiplying together all of the conditional probability distributions.
\begin{equation}\label{eq:1}
  P_{DBN}(x[0],x[1],\cdots,x[T])=P_{B_{0}}\prod_{t=0}^{T-1}P_{B_{\rightarrow}}(x[t+1]|x[t])
\end{equation}
Where $x[t]$ denotes the state of sensor nodes at time $t$.

As mentioned above, it can be seen that the dependencies between sensor nodes in IoT can be described by the structure of Bayesian network. In this paper, based on the Bayesian network model established by sensor nodes, the research on anomaly detection and redundancy elimination for big sensor data in IoT is conducted.
\section{Learning the Structure of Bayesian Network Using Gathered Sensor Data}
Building a specific Bayesian network can be described as finding a suitable structure of network while a training data set $D$ is given. And a Bayesian network is represented as $B=(S,\theta)$ where $S$ is the structure of network (i.e., determining what depends on what) and $\theta$ is a parameter (i.e., the strength of these dependencies)\cite{larranaga2013a}. In order to build the Bayesian network, we divide the sensor data set into several random variables and get transition probability matrix through statistics approach\cite{Song2009Time}. We use a score metrics to get the degree of matching between the training set $D$ and the structure $S$, the probability of structure $S$ given data set $D$ can be written as:
\begin{equation}\label{eq:2}
  P(S|D)=\frac{P(S)P(D|S)}{P(D)}=\frac{P(S)\int_{\theta}P(D|S,\theta)P(\theta|S)\,d\theta}{P(D)}
\end{equation}
Thus, we can depend on the score metrics to search for the best Bayesian network. As we mentioned above, the main point of learning the structure of a network is to get the parent nodes of one specific node. From Eq.(\ref{eq:2}) we can get $P(S|D) \propto P(S)P(D|S)$, so a simplify score metrics is defined as Eq.~(\ref{eq:score}).
\begin{equation}\label{eq:score}Score=\log P(D|S,\theta_{s})\end{equation}
Where $\theta_{s}$ is the estimate optimal parameter which maximizes the likelihood function.
For a dynamic network we give the following definition:
\begin{equation}\label{eq:3}
  \theta_{i,j,k}^{0}= P(X_{i}[0]=k|pa(X_{i}[0])=j)\quad
  \theta_{i,j,k}^{\rightarrow}=P(X_{i}[t]=k|pa(X_{i}[t-1])=j)
\end{equation}

Where $\theta_{i,j,k}^{0}$ is the conditional probability of $X_{i}$ being in its $k^{th}$ value given its parents $pa(X_{i}[0])$ in state j. $\theta_{i,j,k}^{\rightarrow}$ denotes in the transition network the conditional probability of $X_{i}$ in its $k^{th}$ state at time $t$ given its parents $pa(X_{i}[t-1])$ in state $j$. All of those conditional probability can be gotten using statistical methods. Because the initial network describes the dependencies among the nodes at the same time, and the transition network describes the dependencies of the nodes between two temporal slices. So we define a counting rule for initial and transition networks as follows:
\begin{equation}\label{eq:4}
C_{0}(X_{i}[t]=k,pa(X_{i}[t])=j)\!=\!\begin{cases}
1 &X_{i}[t]=k,pa(X_{i}[t])=j\\
0 &otherwise\\
\end{cases}
\end{equation}
According to the counting rule above, it is easy to get the number of specific state appeared in initial and transition network:
\begin{equation}\label{eq:5}
\begin{split}
  N_{i,j,k}^{0} =\sum_{l}C_{0}((X_{i}[0]=k,pa(X_{i}[0])=j);X^{l}) \\
  N_{i,j,k}^{\rightarrow} =\sum_{l}C_{\rightarrow}((X_{i}[t]=k,pa(X_{i}[t-1])=j);X^{l})
\end{split}
\end{equation}
Where $\ell$ denotes the number of training sequences. And in the training data, $N_{i,j,k}$ denotes the number of node $X_{i}$ being in its $k^{th}$ state given its parents in state $j$.

According to the methods mentioned above, we can get the conditional probability in initial network and transition network:
\begin{equation}\label{eq:6}
  \theta_{i,j,k}^{0}=\frac{N_{i,j,k}^{0}}{\sum_{k}N_{i,j,k}^{0}} \quad
  \theta_{i,j,k}^{\rightarrow}=\frac{N_{i,j,k}^{0}}{\sum_{k}N_{i,j,k}^{\rightarrow}}
\end{equation}
Consider the joint probability distribution of DBNs, the likelihood function of a specific training data sets given a possible network structure could be expressed as:
\begin{equation}\label{eq:7}
P(D|S,\theta_{s})=\prod_{i}\prod_{j}\prod_{k}(\theta_{i,j,k}^{0})^{N_{i,j,k}^{0}}\times\prod_{i}\prod_{j}\prod_{k}(\theta_{i,j,k}^{\rightarrow})^{N_{i,j,k}^{\rightarrow}}
\end{equation}
Thus, the score metrics can be gotten
\begin{equation}\label{eq:8} Score = \log P(D|S,\theta_{s})  = \sum_{i}\sum_{j}\sum_{k}N_{i,j,k}^{0}\times\log\theta_{i,j,k}^{0} + \sum_{i}\sum_{j}\sum_{k}N_{i,j,k}^{\rightarrow}\times\log\theta_{i,j,k}^{\rightarrow}\end{equation}

From the score metrics we can learn that the function is formed with two parts. One is the parameters in initial network, the other is the parameters in transition network. Therefore, the structure of initial and transition network can be learned separately. One of the most common used algorithm for learning the structure of Bayesian network is K2 algorithm\cite{Bouchaala2010Improving}. We combine the score metrics which is mentioned above with K2 algorithm, and K2 is like a greedy algorithm which maximizes the score of metrics. Because the structure of Bayesian network is directed acyclic graph (DAG)\cite{Cheng2002Learning}, in order to avoid cyclic graphs in the learned structure of Bayesian network, the K2 algorithm assumes an initial ordering of the nodes such that, if $X_{j}$ proceeds $X_{i}$ in the order, an arc from $X_{j}$ to $X_{i}$ is not allowed. But the disadvantage is that the initial ordering should be based on prior expert knowledge, and in fact it is difficult to get the prior knowledge in practical environment. Thus, in practice, we do not consider the initial ordering. First of all, we use K2 to get the dependencies of each node, then modify the cyclic graph part in the network.

\section{Anomaly Detection for Big Sensor Data in IoT}
Currently, data anomaly detection has already become one of the most popular research directions. But there is few research on anomaly detection for big sensor data. With the characteristics of sensor nodes in the IoT, the relationship of these nodes can be described by Bayesian network. And in this section, the method of anomaly detection for big sensor data base on the two principal statistic models and Bayesian network of sensor nodes is proposed. The anomaly detection algorithm consists of two phases: ``rough" detection stage and ``careful" detection stage. In the ``rough" detection stage, features of the collected big sensor data are extracted according to the two statistic models of the principal component. And then, the anomaly detection for the gathered data streams can be conducted based on the extracted data features. However, in the stage of ``rough" detection, only whether the gathered data stream is abnormal or not can be determined whereas specific anomalous data can not be obtained. Therefore, the ``careful" stage, when exact abnormal data can be obtained according to the Bayesian network learned by big sensor data, is necessary.
\subsection{Sensor Data Anomaly Detection Based on Squared Prediction Error (SPE) and Hotelling's $T^{2}$ Statistics}
The SPE statistic mainly describes the degree of samples collected at current time deviation from the principal component. Based on the feature extraction of the sensor data sets, if a data stream collected at a certain moment is deviated too much from the principal component characteristics, it means the data stream may be abnormal.

The sensor data collected by multiple sensor nodes can be expressed as a data matrix $X_{m\times n}$, where $m$ is the number of samples, $n$ is the number of sensor nodes $(m>n)$. In order to eliminate the impact of individual data on the whole samples, the data matrix $X_{m\times n}$ needs to be standardized using Eq. (\ref{eq3}).
\begin{equation}\label{eq3}
  \overline{X}=[X-I_{n}v^{T}]D^{-1/2}
\end{equation}
Where $I_{n}$ is a $n\times n$ identity matrix, $v=[v_{1},v_{2}\cdots v_{n}]^{T}$ is the vector of the mean value of each sensor node, $D=diag(\sigma_{1}^{2},\sigma_{2}^{2}\cdots \sigma_{n}^{2})$ denotes a diagonal matrix and the diagonal values are sample variance of each node.

The core idea of data anomaly detection based on SPE statistic is to reconstruct the data stream collected by multi-sensor nodes at the current time according to the features extracted from the training data sets, and then according to the reconstructed error judge whether the data stream is abnormal. Therefore, with principal component analysis we can get the eigenvalues $\lambda_{i}$ and feature vector $p_{i}$ of standardization matrix $\overline{X}$, where $i=1,2\cdots n$. Based on the method of selecting the number of principal elements, the $k(k <n)$ eigenvectors are selected to reconstruct the standardization matrix as follows:
\begin{equation}\label{eq4}
  \widetilde{X}\approx S_{k}P_{k}^{T}=s_{1}p_{1}^{T}+s_{2}p_{2}^{T}+\cdots+s_{k}p_{k}^{T}
\end{equation}
Where $S_{k}=[s_{1},s_{2}\cdots s_{k}]$ is a matrix composed of the principal component score vectors, $s_{k}=\overline{X}p_{k}, k=1,2\cdots n$, and $P_{k}$ is the feature matrix of $\overline{X}$.

Thus, we can get the reconstruction error of the matrix $E=\overline{X}-\widetilde{X}$. Then the squared prediction error of the data samples collected by the multiple sensor nodes at time $i$ can be expressed as follows:
\begin{equation}\label{eq5}
  SPE(i)=\Sigma_{j=1}^{n}(\overline{X}_{ij}-\widetilde{X}_{ij})^{2},i=1,2\cdots m
\end{equation}
Where $\overline{X}_{ij}$ is the standardization value of $j^{th}$ sensor node collecting data at the $i^{th}$ time, $\widetilde{X}_{ij}$ is the reconstruction data.

For convenience, we use $Q$ (Eq. \ref{eq6}) to denote the statistic which can express the squared prediction error of the data sets collected by sensor nodes at the $i^{th}$ time:
\begin{equation}\label{eq6}
  Q(i)=e_{i}e_{i}^{T}=\overline{X}_{i}(I-P_{k}P_{k}^{T})\overline{X}_{i}^{T}
\end{equation}
Where $e_{i}$ is the $i^{th}$ row of reconstruction error matrix $E$, $P_{k}$ is the feature matrix which is composed of selected $k$ principal component eigenvectors. $I$ is the $n\times n$ identity matrix. $\overline{X}_{i}$ is the standardization value of the sensor data collected at time $i$.

It can be seen that the value of $Q$ statistic is scalar at a specific time. As mentioned above, it describes the degree of samples collected at current time deviation from principal component. And the degree of deviation can be determined by setting the threshold of the $Q$ statistic. When the test level is $\alpha$, the threshold of the $Q$ statistic can be given by Eq.~(\ref{eq7}):
\begin{equation}\label{eq7}
   Q_{\alpha} =\theta_{1}|\frac{C_{\alpha}\sqrt{2\theta_{2}h_{0}^{2}}}{\theta_{1}}+\frac{\theta_{2}h_{0}(h_{0}-1)}{\theta_{1}^{2}}+1|^{\frac{1}{h_{0}}}\quad
   \theta_{i} =\sum_{j=k+1}^{n}\lambda_{j}^{i},(i=1,2,3)\quad
    h_{0}=1-\frac{2\theta_{1}\theta_{3}}{3\theta_{2}^{2}}
\end{equation}

Where $C_{\alpha}$ is the critical value of the normal distribution at the test level $\alpha$. $\lambda_{j}$ is the eigenvalue of the standardization data matrix. $K$ is the number of selected principal components. $N$ is the number of sensor nodes. According to Eq.~(\ref{eq6}) and Eq.~(\ref{eq7}), we can get the value and threshold of $Q$ statistic. If the value of the $Q$ statistic is greater than the threshold, it indicates that the test data stream is anomalous.

$T^{2}$ statistic is a commonly used multivariate test method. It reflects the change of projected data on the principal component subspace. $T_{i}^{2}$ reflects the degree of the trend and amplitude value deviation from the principal component model for the sample sensor data gathered at time $i$.
For the defined data matrix $X_{m\times n}$, $X_{i}$ denotes the data stream collected at time $i$ where $i=1,\cdots m$ and the value of $T_{i}^{2}$ can be expressed as follows:
\begin{equation}\label{eq11}
  T_{i}^{2}=t_{i}\lambda^{-1}t_{i}^{T}=\overline{X}_{i}P_{k}\lambda^{-1}P_{k}^{T}\overline{X}_{i}^{T}
\end{equation}
Where $\lambda$ is the $k\times k$-dimensional diagonal matrix formed by the first $k$ eigenvalues selected from the principal feature. $P_{k}$ is the matrix of eigenvectors corresponding to the selected $k$ eigenvalues, $t_{i}$ represents the score vector in $k$ principal directions for the data collected by each sensor node at $i^{th}$ time.

The value of $T^{2}$ and $\frac{(m-1)n}{(m-n)}F_{n,(m-n)}$ are identically distributed, where $F_{n,m-n}$ denotes an F-distributed random variable with degrees of freedom $n$ and $m-n$. And the threshold of the $T^{2}$ statistic can be given by Eq.~S(\ref{eq12}):
\begin{equation}\label{eq12}
  T_{k,n,\alpha}^{2}=\frac{k(m-1)}{m-k}F_{k,m-1,\alpha}
\end{equation}
Where $\alpha$ is the significant level, $n$ is the number of sensor nodes, $m$ is the number of samples.
The value of $\alpha$ can determine the boundaries of anomaly detection. By setting $\alpha=0.05$, we can get the warning boundary. When $\alpha=0.01$, the alarm boundary can be determined.

A number of variables being monitored at the same time can be achieved through the $T^{2}$ statistic. For a specific time, if the $T^{2}$ statistic value of a sensor data stream is greater than the threshold, it means the collected data stream and the training data matrix $X$ do not obey the same distribution and it is regarded as an abnormal data stream.

\subsection{Sensor Data Anomaly Detection Algorithm Based on Principal Statistic Analysis and Bayesian Network}
 \begin{figure}[!t]
	\centering
	\includegraphics[height=3cm,width=9cm]{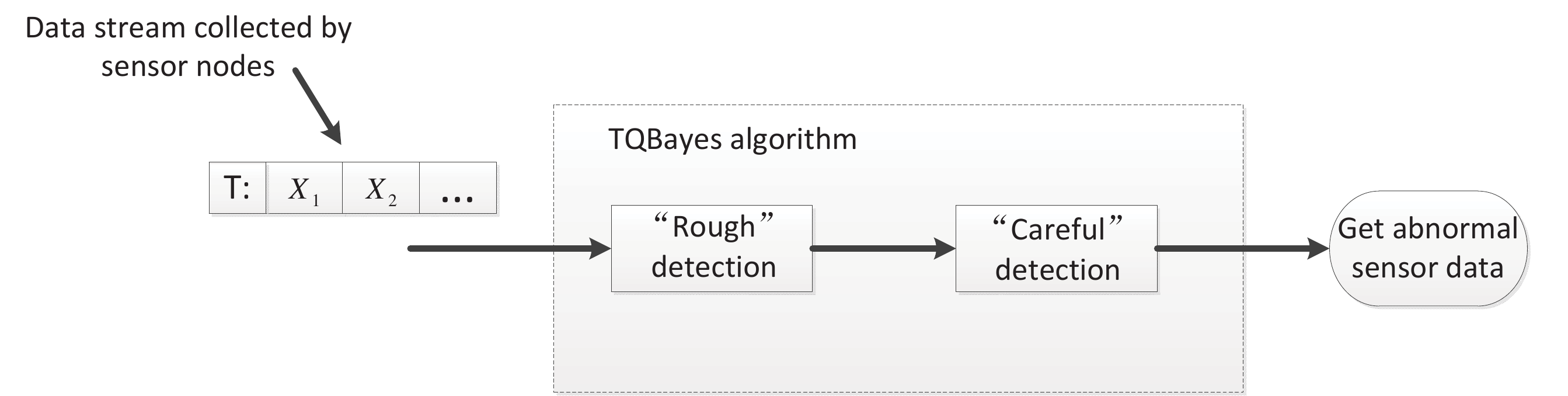}
	\caption{The process of anomaly detection of TQBayes algorithm}
	\label{fig:fig4}
\end{figure}
\renewcommand{\algorithmicrequire}{\textbf{Input:}}
\renewcommand{\algorithmicensure}{\textbf{Output:}}
\begin{algorithm*}[!t] 
\caption{Sensor Data Anomaly Detection Algorithm Based on Principal Statistic Analysis and Bayesian Network} 
\label{alg:TQBayes} 
\begin{algorithmic}[1] 
\REQUIRE
\begin{enumerate}
\item[]
\item[-]Training data sets $X_{p\times n}$
\item[-]Testing data sets $X_{m\times n}$
\end{enumerate}
\ENSURE
Abnormal data points 
\STATE Standardize the training data set and get matrix $\overline{X}_{p\times n}$ \STATE With principal component analysis for $\overline{X}_{p\times n}$ the eigenvalues and eigenvectors can be obtained.
\STATE According to the cumulative contribution rate of eigenvalues, the number of principal component $k$ can also be determined.
\STATE According to Eqs.~(\ref{eq7}),(\ref{eq12}) the threshold $Q$ and $T^{2}$ of the two principal components can be obtained.
\STATE According to the training data set, we can obtain the Bayesian networks which describe the dependencies among the sensor nodes.
\FOR{$t=1$ to $m$}
\STATE Standardize the data blocks of $n$ sensor nodes acquired at time $t$ and obtain vector $\overline{X}_{1\times n}$.
\STATE According to Eqs.~(\ref{eq6}),(\ref{eq11}) get the two statistic values $Q_{t}$ and $T_{t}^{2}$.
\IF{$(Q_{t}>Q || T_{t}^{2}>T^{2})$}
\STATE For a suspicious node get its parent nodes from the structure of Bayesian network
\STATE The method of Naive Bayesian classifier (Eq.(\ref{eq13})) is used to find out which node is abnormal.
\ENDIF
\ENDFOR
\end{algorithmic}
\end{algorithm*}

The aforementioned method for detecting anomalies is based on principal statistic analysis, and we named it ``rough" anomaly detection method. This method can only determine whether there is abnormal in the gathered data stream at a specific time. Yet it can not determine which node is abnormal. In order to solve this problem, in this section, we propose to establish the Bayesian network which can reflect the relationship of sensor nodes to determine the abnormal nodes.

According to structure learning method of Bayesian network introduced in Section 3, we can easily get the dependency of sensor nodes between two time slots through the training data set. And we clearly divide the gathered sensor data into several states. If each state of a sensor node is considered to be one category, then the problem of inferring the state of current node from the state of its parents can be seen as a classification problem given the state of parent nodes. So after the establishment of the Bayesian network, the state inference can use Naive Bayes classifier to solve the problem.

We use the Naive Bayes classifier to infer the state of a node at a specific time, and the state of its parent nodes can be regarded as one feature for state inference. Thus, the state inference based on Naive Bayesian classifier can be express as Eq.~(\ref{eq13}). After ``rough" detection stage, for the detected anomaly data stream if the inference state of specific node is different from its original state, it means the data gathered by this node is abnormal at current time.
\begin{equation}\label{eq13}
  P(X_{i,t}|pa_{1,(t-1)},pa_{2,(t-1)}\cdots pa_{2,(t-1)})
     = \underset{j=1}{\overset{n}{\prod}}P(X_{i,t}|pa_{j,(t-1)})P(X_{it})
\end{equation}
Where $X_{i,t}$ denotes the state of node $X_{i}$ at time $t$, $pa_{j,(t-1)}$ denotes the state of parent node $pa_{j}$ at time $t-1$.

Alg.~\ref{alg:TQBayes} shows the sensor data anomaly detection algorithm based on principal statistic analysis and Bayesian network. For convenience, we named the algorithm $TQBayes$. Fig.~\ref{fig:fig4} shows the process of sensor data anomaly detection based on the proposed $TQBayes$ method. And the process is composed of two stages. In the first stage, a rapid rough detection method based on principal statistic analysis is proposed to identify the data stream that may be abnormal. Secondly, for further detection, through the method of Naive Bayesian classifier based Bayesian network of sensor nodes, we can find out the correct node that generates abnormal data. The combination of these two methods can improve the efficiency of abnormal detection under the premise of ensuring the accuracy of the algorithm.
\section{Redundancy Elimination of Big Sensor Data in IoT}
In a sensor network, there are many factors which cause data redundancy. For example, where the gap among each node is close, the type of collecting data is similar. Redundant data not only waste the storage space but also exert harmful influence on data feature extraction. In this part, we mainly focus on the methods of redundancy elimination directly from the perspective of gathered sensor data. Two methods are proposed for static and dynamic sensor data redundancy elimination separatively.
\subsection{Static Sensor Data Redundancy Detection}
The sensor data that stored in database is regarded as static data. We propose a static data redundancy detection algorithm (SDRDA) by building the SBN of the sensor nodes. According to the dependencies reflect in the SBN, the inference of redundant node can be figured out.

\begin{figure}[!htp]
\centering
\subfigure[]{
\label{fig:sbn1}
\includegraphics[height=3cm,width=3cm,scale=0.5]{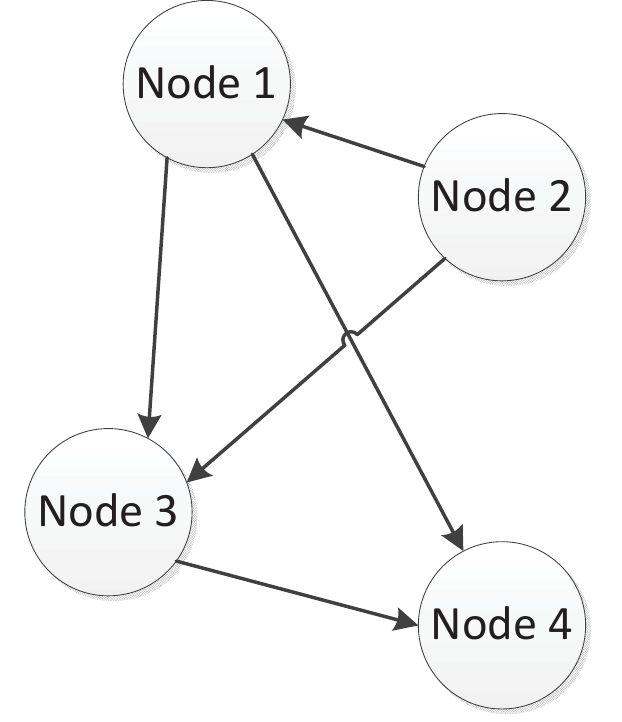}}
\subfigure[]{
\label{fig:matrix}
\includegraphics[height=2cm,width=3cm,scale=0.5]{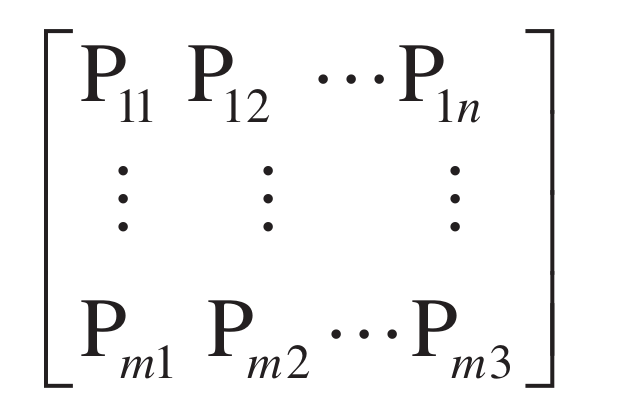}}
\caption{An example of Bayesian network with four sensor nodes and transition probability matrix}
\end{figure}

Fig.~\ref{fig:sbn1} shows an example of a Bayesian network structure for a problem with four nodes. We can get the parent nodes of one specific node in the network. Fig.~\ref{fig:matrix} shows the transition probability matrix between current node and its dependent nodes. The row of the matrix denotes the number of state of parent nodes, and the column of the matrix denotes the number of state of current node. For each row $\sum_{i=1}^{n}P_{ki}=1$,where $k=1\cdots m$.

\renewcommand{\algorithmicrequire}{\textbf{Input:}}
\renewcommand{\algorithmicensure}{\textbf{Output:}}
\begin{algorithm*}[!t] 
\caption{Static Sensor Data Redundancy Detection Algorithm} 
\label{alg2} 
\begin{algorithmic}[1] 
\REQUIRE
\begin{enumerate}
\item[]
\item[-] A set of nodes $X=\{X_{1},X_{2}\cdots,X_{N}\}$             
\item[-] The Bayesian Network of N nodes
\end{enumerate}
\ENSURE  A printout of node redundancy 
\FOR{$i=1$ to $N$}
\STATE isRedundancy=false
\STATE set $ Pa_{i} $ to empty:$Pa_{i}=\phi$
\STATE initialize the transition probability between $X_{i}$ and its parents\\
 $transMatrix=\phi$
\STATE $Pa_{i}=findTheParent(X_{i})$
\STATE $transMatrix=createTransMatrix(X_{i},Pa_{i})$
\IF{$\sum_{h=1}^{H}max(P(X_{t}^{i}=s_{ik}|Pa(X_{t}^{i})=s_{ih}^{pa}))\longrightarrow H$}
\STATE $isRedundancy=true$
\ENDIF
\ENDFOR
\end{algorithmic}
\end{algorithm*}
With the dependencies of each node, a specific node can form a subnet with its parent nodes. And according to the data sets those sensor nodes collected, we can get a state transition probability matrix of each subnet through statistical methods. If the conditional probability of $X_{t}^{i}$ being in state $s_{ik}$ approaches to 1, given that its parent in state $s_{i}^{pa}$, it indicates that we can infer the state of $X_{t}^{i}$ by its parents node. And define it as:
\begin{equation}P(X_{t}^{i}=s_{ik}|Pa(X_{t}^{i})=s_{i}^{pa})\longrightarrow1\end{equation}
Thus, if all of the state of parent nodes can inference the state of current node, we regard current node as redundant node. The conditional probability can be defined as:
\begin{equation}\sum_{h=1}^{H}max(P(X_{t}^{i}=s_{ik}|Pa(X_{t}^{i})=s_{ih}^{pa}))\longrightarrow H\end{equation}
Where $X_{t}^{i}$ denotes node $i$ at time $t$, $s_{ik}$ denotes node $i$ in its $k^{th}$ state, $Pa(X_{t}^{i})$ denotes the parent nodes of node $i$ at time $t$, $s_{ih}^{pa}$ denotes the parent nodes in its $h^{th}$ state, $H$ denotes the number of the state of parent nodes.

Alg. \ref{alg2} shows the proposed algorithm for static sensor data redundancy detection. With this algorithm the redundant data in a collected sensor data sets can be dug out.
\subsection{Real-time Sensor Data Redundancy Detection}
\renewcommand{\algorithmicrequire}{\textbf{Input:}}
\renewcommand{\algorithmicensure}{\textbf{Output:}}
\begin{algorithm*}[!t] 
\caption{Real-time Sensor Data Redundancy Detection} 
\label{alg3} 
\begin{algorithmic}[1] 
\REQUIRE \begin{enumerate}
\item[]
\item[-]A set of nodes $X=\{X_{1},X_{2}\cdots,X_{N}\}$             
\item[-]The transition network for all node
\item[-]The transition probability matrix for each node
\item[-]A data set which is collected at $t-1$ for all node
\end{enumerate}
\ENSURE a printout of the state of each node at time $t$. (waking/sleeping)
\FOR{$i=1$ to $N$}
\STATE initialize the sensor nodes in state ``waking":$Sensor\_state=waking$
\STATE set $ Pa_{i} $ to empty:
 $ Pa_{i}=\phi $
 \STATE Find the parent nodes of $X_{i}$ according to the learned Bayesian Network\\
$Pa_{i}=findTheParent(X_{i})$
\STATE Get the transition probability between $X_{i}$ and its parents\\
 $transMatrix=createTransMatrix(X_{i},Pa_{i})$
\STATE Get the states probability table of the parent of $X_{i}$ at  $t-1$, elements in this table like $P(X=states_{i}),i=1,2\cdots$
\FOR{$s=1$ to $n$}
\STATE $P(X_{s}|e^{pa})=\sum_{i,j\cdots
k}P(X_{s}|pa_{1i},pa_{2j},\cdots,pa_{|pa|k})\times\prod_{m=1}^{|pa|}P(pa_{m}|e_{pa_{m}})$
\ENDFOR
\STATE Normalize the condition probability
\IF{$max(P(X_{s}|e^{pa}))\rightarrow1$}
\STATE $Sensor\_state=sleeping$
\ENDIF
\ENDFOR
\end{algorithmic}
\end{algorithm*}
In previous section, we have mentioned an algorithm for static sensor data redundancy detection. As we know, redundant data is collected by redundant node. Whether there is a way that we can detect the redundant node while it is working?  If a specific node is detected as redundant node at time $t$, and the node can be sleeping at this time. On one hand it is better for reducing data redundancy, on the other hand some nodes in the sensor network may not be working all the time. A DBN is an extension of  SBN to temporal domain, in which conditional dependencies are modeled between random variables both within and across time slots\cite{Wang2011Time}. Thus, it is suitable for solving real-time inference problem. According to the characteristic of DBN, in this section we will post a method to build a DBN structure for a working sensor network.

\begin{figure}
\begin{minipage}[t]{0.5\linewidth}
\centering 
\includegraphics[height=4cm,width=8cm,scale=0.5]{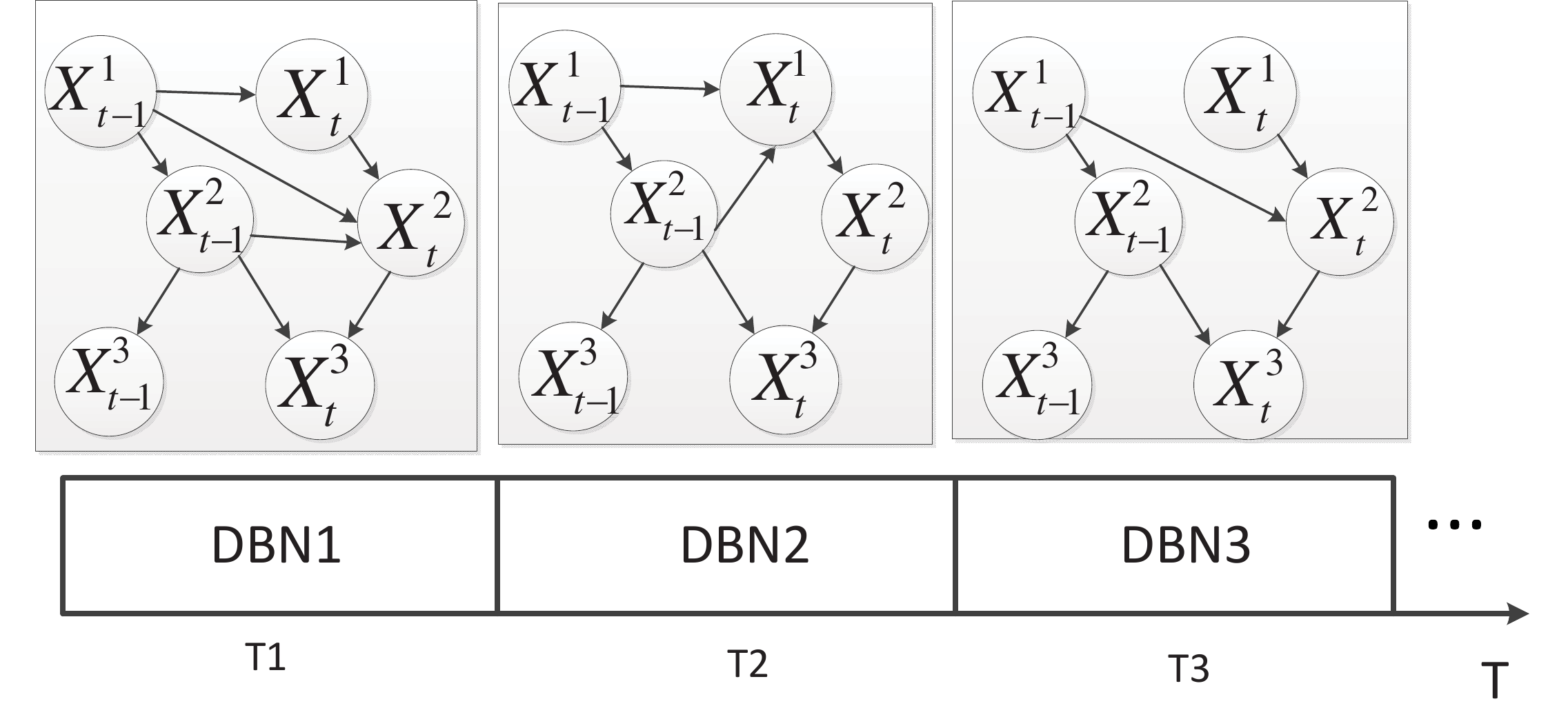}
\caption{The model of variable structure for DBN} 
\label{fig:model of dbn}
\end{minipage}
\begin{minipage}[t]{0.5\linewidth}
\centering 
\includegraphics[height=4cm,width=8cm,scale=0.5]{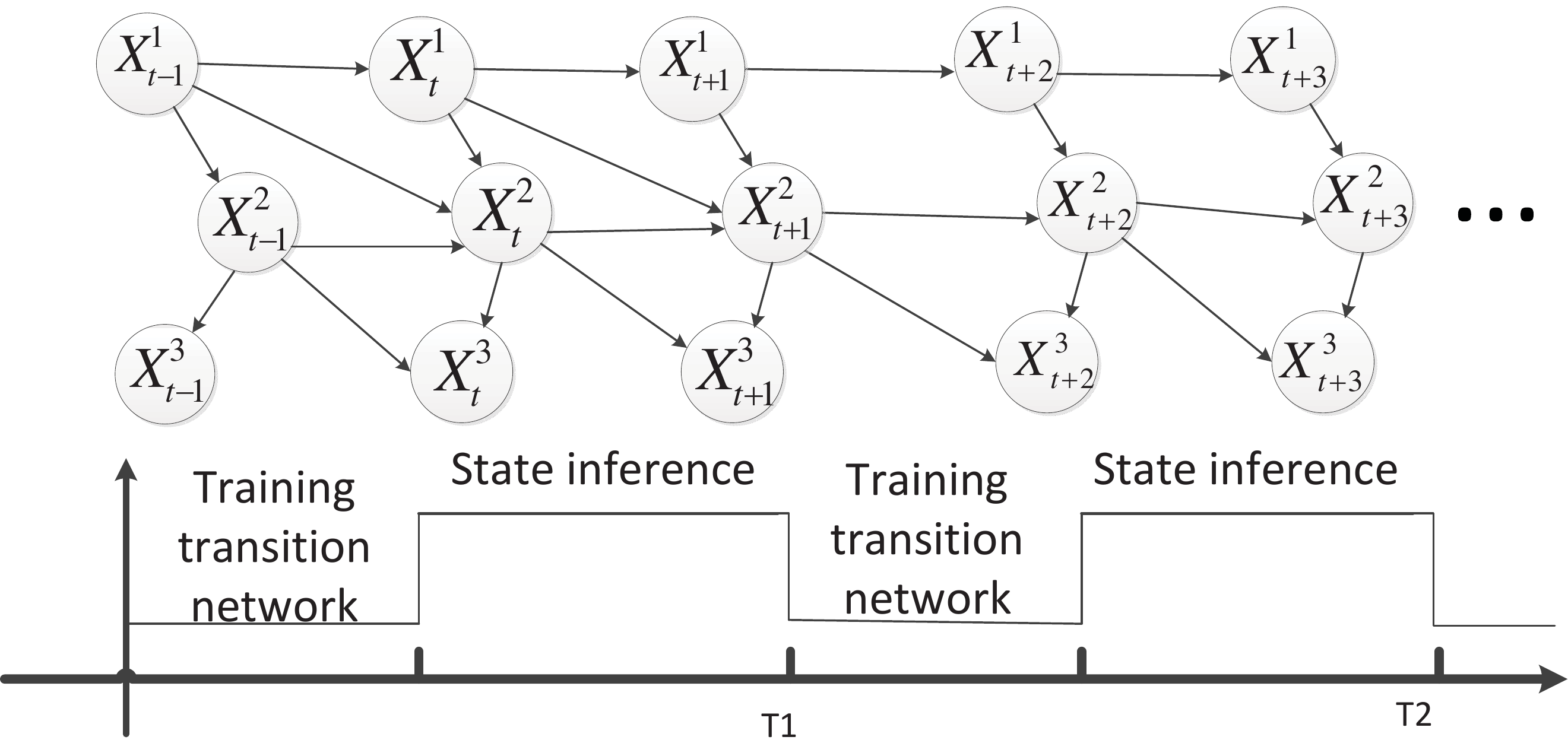}
\caption{The process of real-time redundancy detection} 
\label{fig:process redundancy}
\end{minipage}
\end{figure}

The varying dependencies of each node in DBN reflects the real-time characteristic of a sensor network. The main point of real-time data redundancy detection is that the state of a specific node at time $t$ can be inferred by its dependent node at time $t-1$. So, first of all we should build the real-time dependencies network for the sensor nodes. The dependencies in a sensor network will not change sharply in limit time\cite{Murphy2002Dynamic}. Fig.~\ref{fig:model of dbn} shows a model of variable structure for DBN. We unroll the DBN in $\{T_{1},T_{2}\cdots,T_{n}\}$ slices. The transition structure in each slice is invariable.
Fig.~\ref{fig:process redundancy} shows that we split each time slice into two parts. In the former part all of the nodes are in working state, and according to the data sets which are collected by the nodes in former part, we can learn the structure of Bayesian network in current time slice. As we mentioned above, the structure of Bayesian network will not change sharply within a limited time. Thus, in the second part of the time slice, we use the structure which is trained in former part to predict the working state of each node. Base on this mechanism, the sensor network is in a circle of collecting data, learning transition network, and working state inference.

Bayesian inference in SBN can be extended to DBN~\cite{Murphy2002Dynamic}, the difference between them is that in SBN the parent node at current time is regarded as evidence and in DBN the parent node at previous time is regarded as evidence. DBN mainly focus on the dependencies across two time slots~\cite{Robinson2008Non}. Suppose the state sequence of node $X$ is $\{X_{1},X_{2}\cdots,X_{n}\}$, if the conditional probability of node $X$ in a specific states is approaches to 1, given that the state of its parent nodes at previous time, $P(X_{i}|Pa_{1},Pa_{2},Pa_{3})\rightarrow1$, it denotes that the state of current node $X$ can be inferred by its parent nodes at previous time. So, if the node $X$ still stays in working, the data collected by it can be regarded as redundant data, and node $X$ should be sleeping at this time. Node $X$ in any states can be described as the confidence level of node $X$ given that the state of its parent nodes as evidence. The inference process is as follows:
 \begin{eqnarray} P(X|e^{pa}) &=& P(X|e_{pa_{1}},\cdots,e_{pa_{i}},\cdots,e_{pa_{|pa|}}) \nonumber \\ &=& \sum_{i,j,\cdots k}P(X|pa_{1i},pa_{2j},\cdots,pa_{|pa|k})
 P(pa_{1i},pa_{2j},\cdots,pa_{|pa|k}|e_{pa_{1}},e_{pa_{2}}\cdots e_{pa_{|pa|}})\nonumber \\ &=&
 \sum_{i,j\cdots k}P(X|pa_{1i},pa_{2j},\cdots,pa_{|pa|k})P(pa_{1i}|e_{pa_{1}})\cdots P(pa_{|pa|k}|e_{pa_{|pa|}})\end{eqnarray}
Where $pa_{i}$ denotes $i^{th}$ parent node. $e_{pa_{i}}$ denotes the probability of the state of parent node. $|pa|$ denotes the number of parent nodes. $pa_{mn}$ denotes the value of parent node $pa_{i}$ in state $n$.
Thus, according to the evidence of parent nodes, the inference of current node in a specific state is defined as:
\begin{small}
\begin{equation}\label{eq:9}
P(X_{s}|e^{pa})\!=\!\sum_{i,j\cdots k}P(X_{s}|pa_{1i},pa_{2j},\cdots,pa_{|pa|k})\prod_{m=1}^{|pa|}P(pa_{m}|e_{pa_{m}})\end{equation}
\end{small}Where $s$ denotes the state of node $X$.

From Eq.~(\ref{eq:9}) we can learn that the probability of current node in a specific state is the sum of the prior probability of the parent nodes in all states. We can get the prior probability of the states of parent nodes in previous time through training data sets. By Putting the values of transition and state probability into Eq.~(\ref{eq:9}), we can infer the specific state of current node. And the algorithm of real-time sensor data redundancy detection is shown above.

\section{Experimental Results and Discussion}
In this section, several experiments are conducted to validate the feasibility of the proposed framework on sensor data preprocessing. In the first part, we will make brief introduction about the test dataset. Based on the method described in Section 3, the result of learning the structure of Bayesian network for sensor nodes is presented in the second part. And then, the experimental results of the algorithms proposed for sensor data anomaly and redundancy detection are presented in the third and fourth part.
\subsection{The Result of Building the Structure of Bayesian Network from Gathered Sensor Data}
\begin{figure}
\begin{minipage}[t]{0.5\linewidth}
\centering 
\includegraphics[height=4cm,width=4cm,scale=0.5]{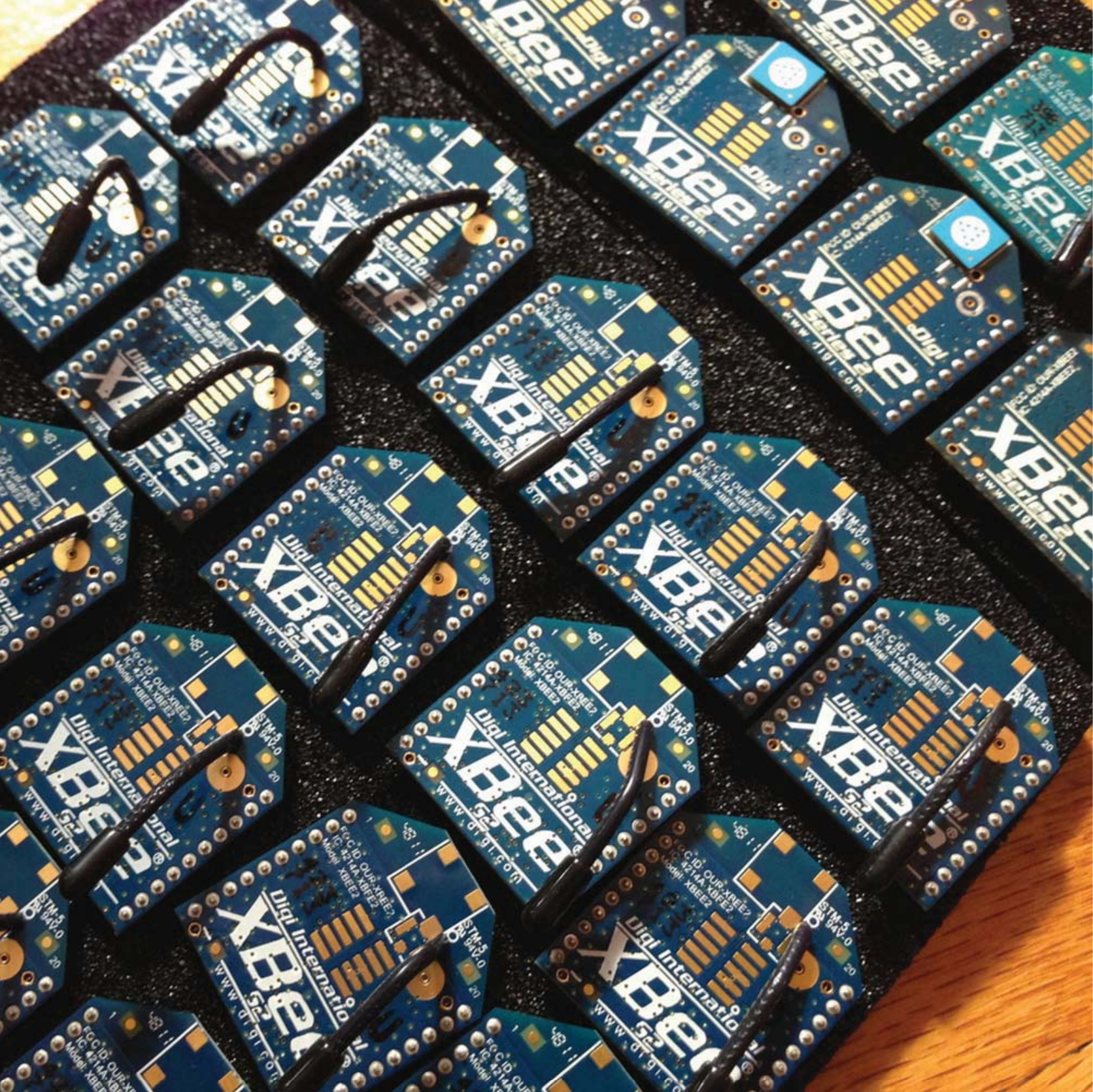}
\end{minipage}
\begin{minipage}[t]{0.5\linewidth}
\centering 
\includegraphics[height=4cm,width=4cm,scale=0.5]{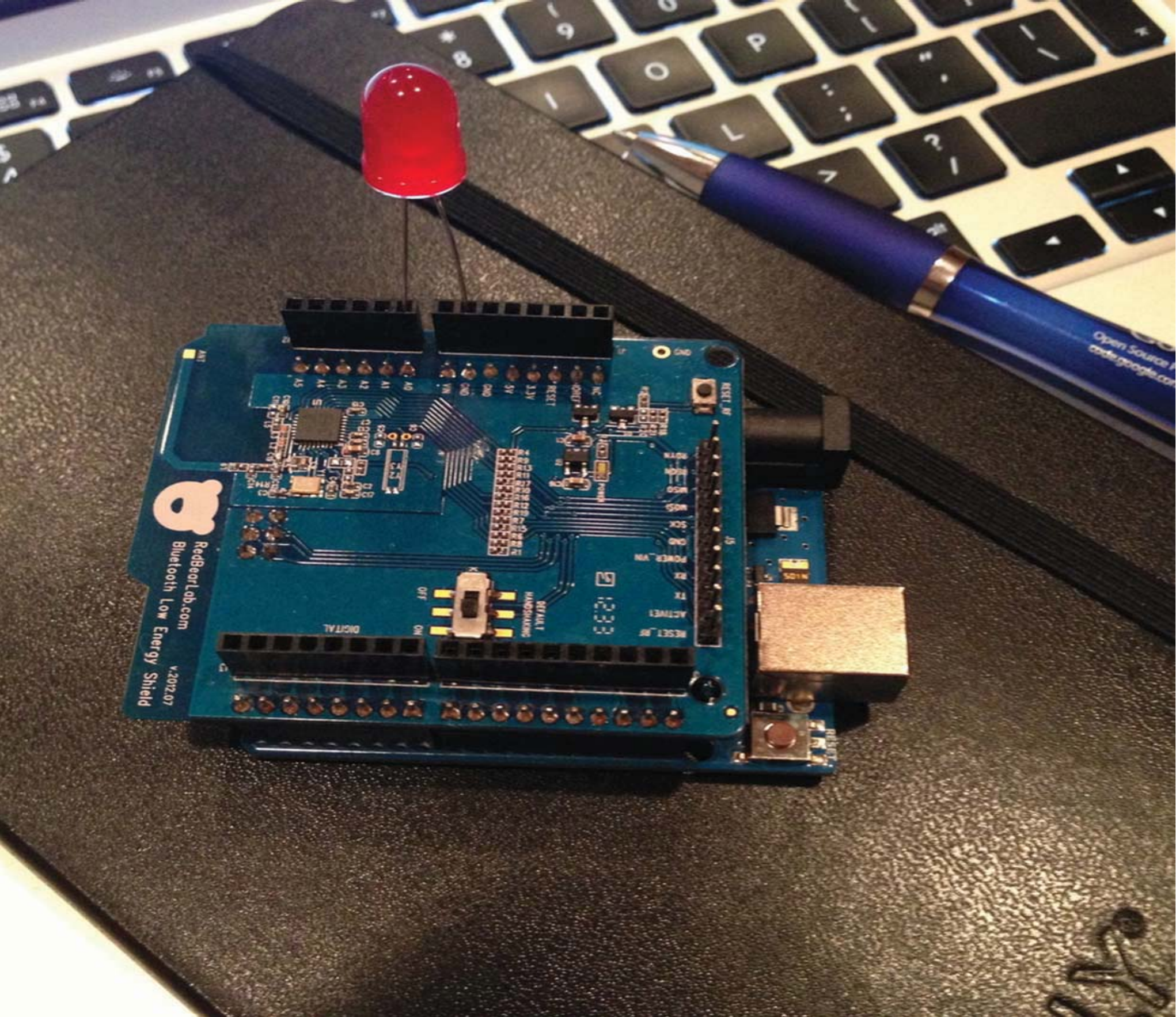}
\end{minipage}
\caption{The Sensor motes\cite{2012Data}}
\label{fig:sensor}
\end{figure}

Our test dataset comes from 40 sensor motes (as shown in Fig.~\ref{fig:sensor}) that are composed of Arduino Leonardo boards, XBee radios, and a handful of off-the-shelf parts, including a PIR motion detector, a temperature and humidity sensor, and an electret microphone amplifier. These motes were distributed around the conference venue, and reported back during the conference. The data were made publicly available online~\cite{2012Data}. These sensor nodes collected temperature, humidity, and microphone values once every one minute.
\begin{figure}[!t]
\centering
\subfigure[The dependencies of temperature sensor nodes]{
\label{fig:temp}
\includegraphics[height=4cm,width=4cm,scale=0.5]{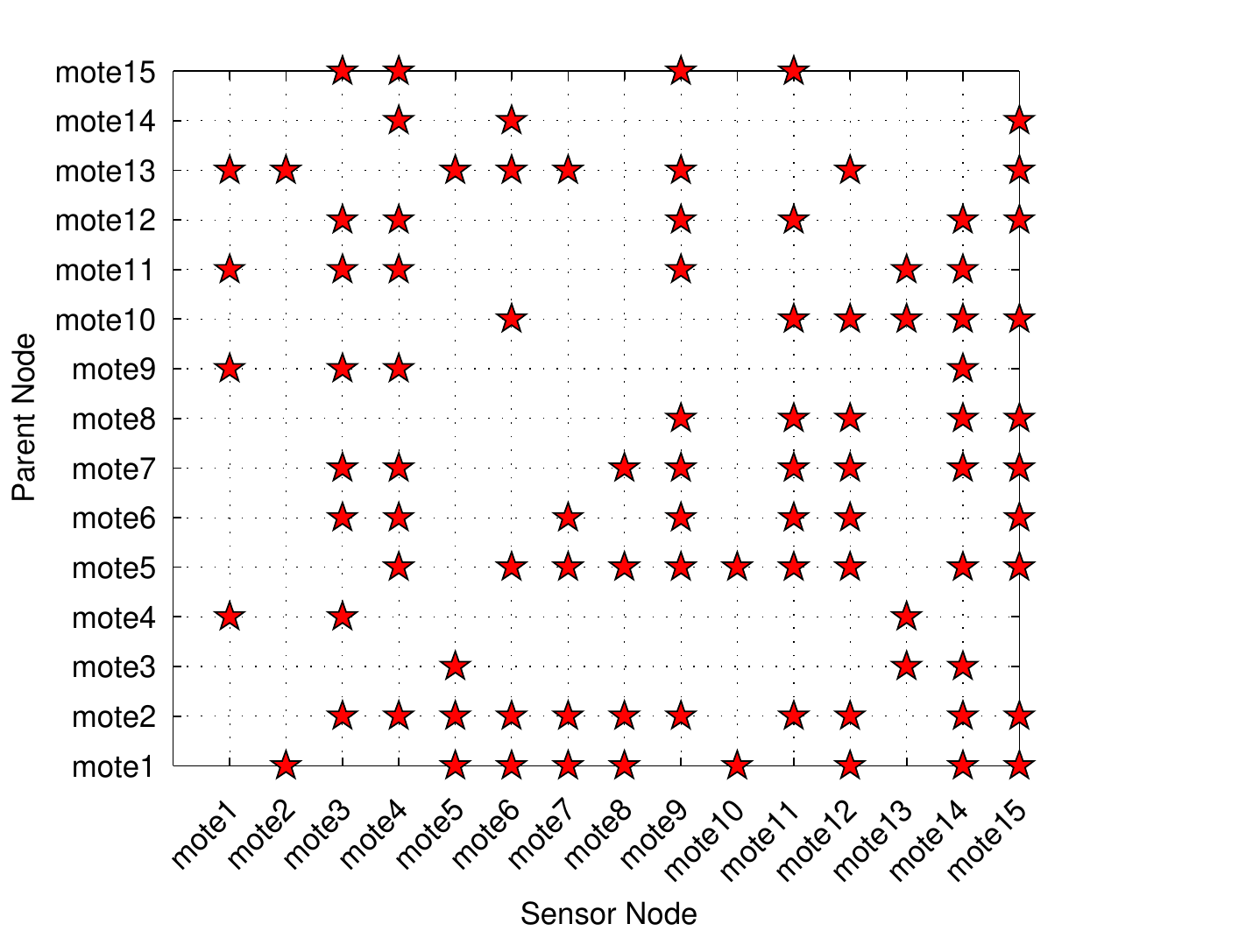}}
\subfigure[The dependencies of humidity sensor nodes]{
\label{fig:hum}
\includegraphics[height=4cm,width=4cm,scale=0.5]{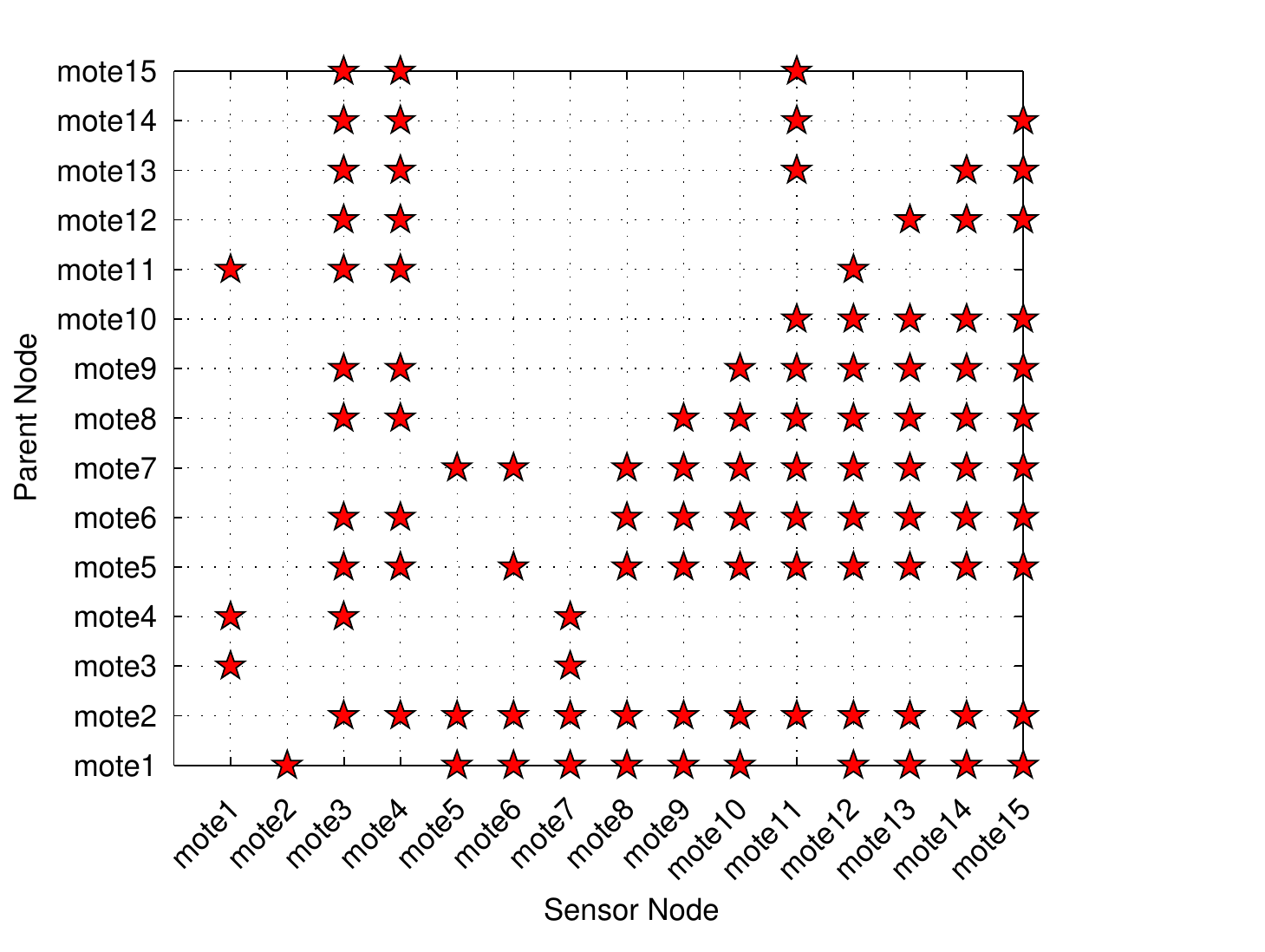}}
\subfigure[The dependencies of microphone sensor nodes]{
\label{fig:mic}
\includegraphics[height=4cm,width=4cm,scale=0.5]{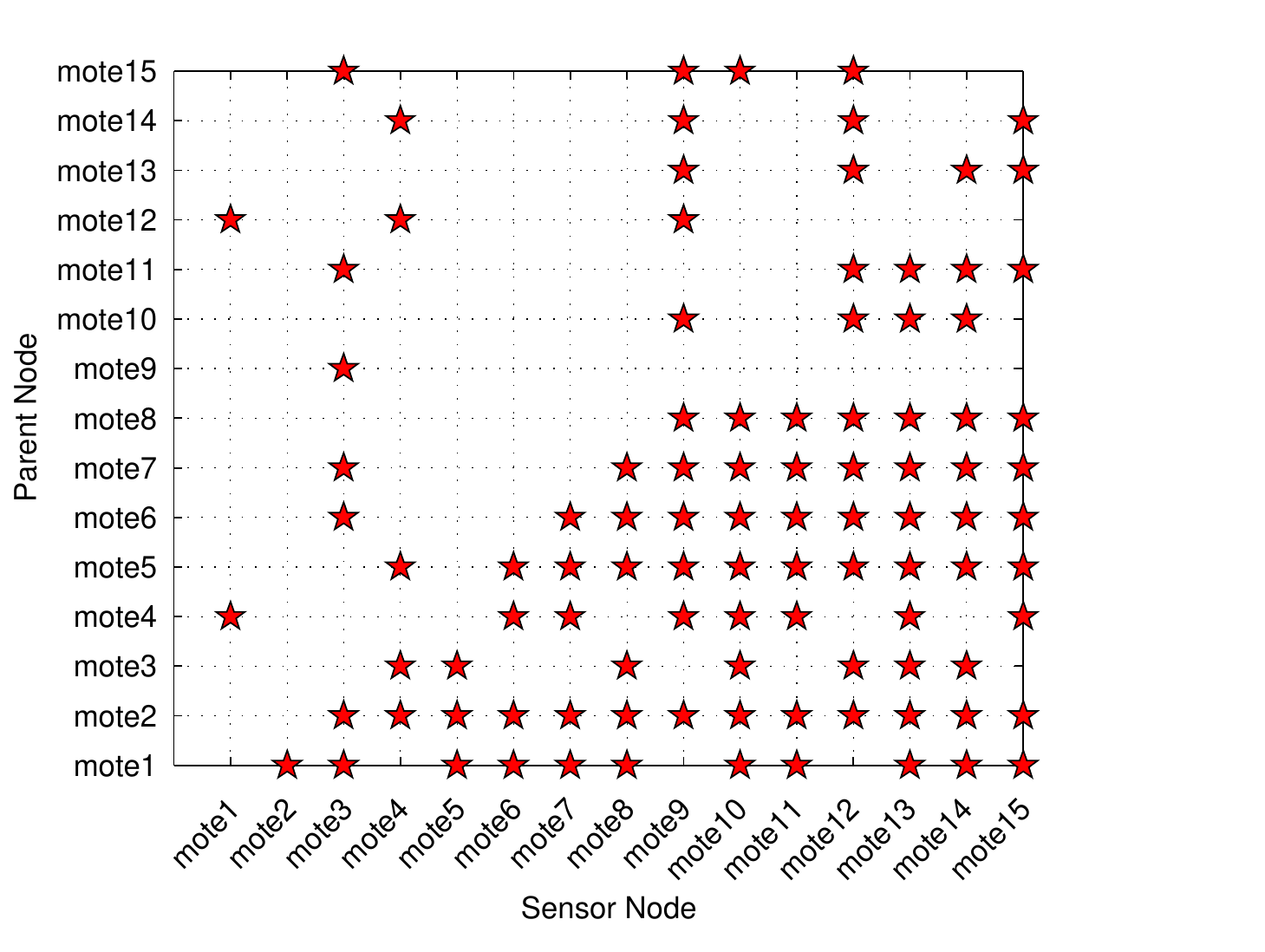}}
\caption{The dependencies of each sensor node in the gathered temperature, humidity, and microphone sensor datasets.}
\label{fig:dependencies}
\end{figure}
\begin{figure*}[!t]
	\centering
     \subfigure[The detection result of adding 1\% simulated errors based on Q statistic]{
      \label{fig:Q1}
     \includegraphics[height=4.5cm,width=5cm]{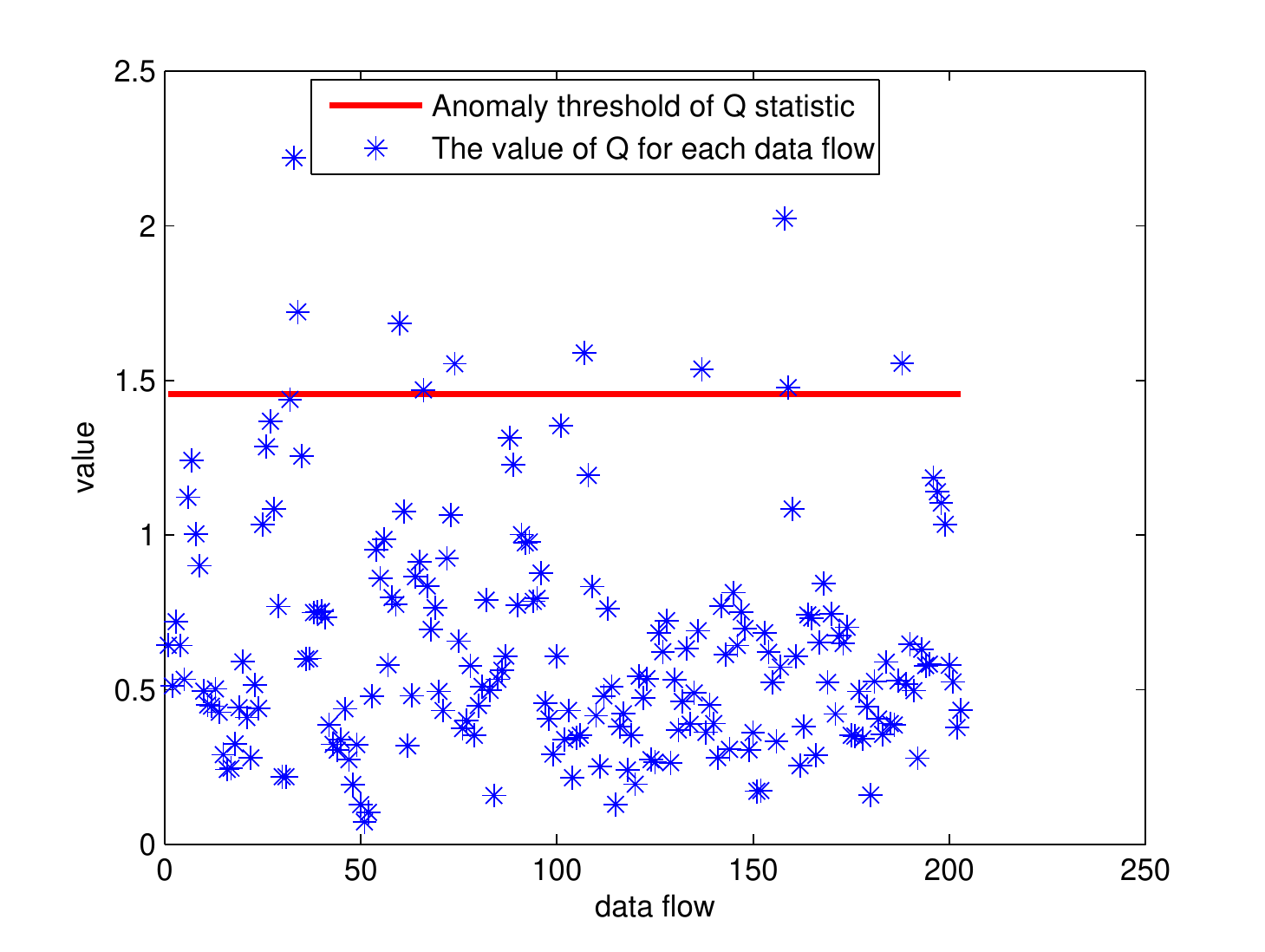}}
     \subfigure[The detection result of adding 5\% simulated errors based on Q statistic]{
      \label{fig:Q2}
     \includegraphics[height=4.5cm,width=5cm]{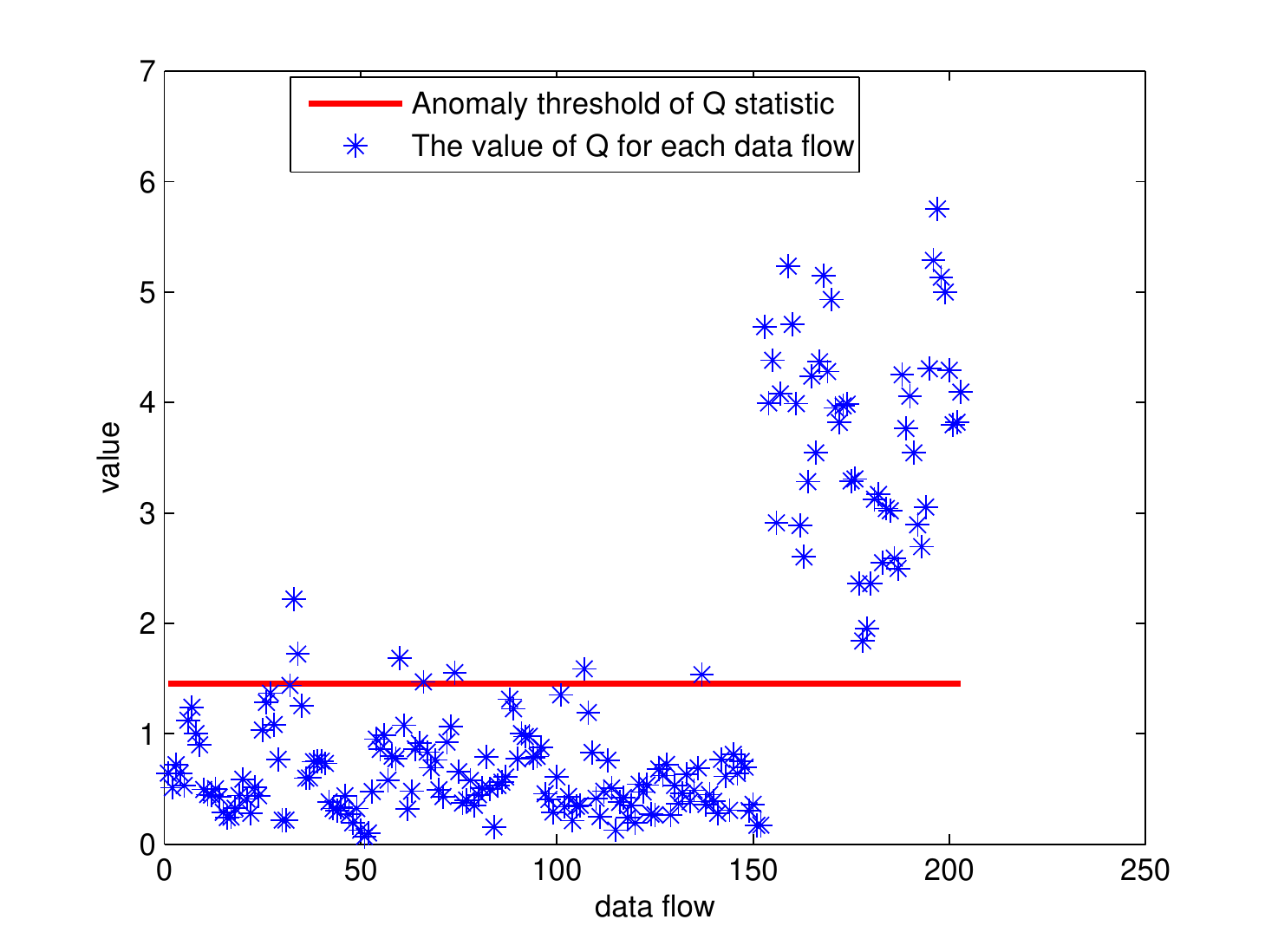}}
     \subfigure[The detection result of adding 10\% simulated errors based on Q statistic]{
      \label{fig:Q3}
     \includegraphics[height=4.5cm,width=5cm]{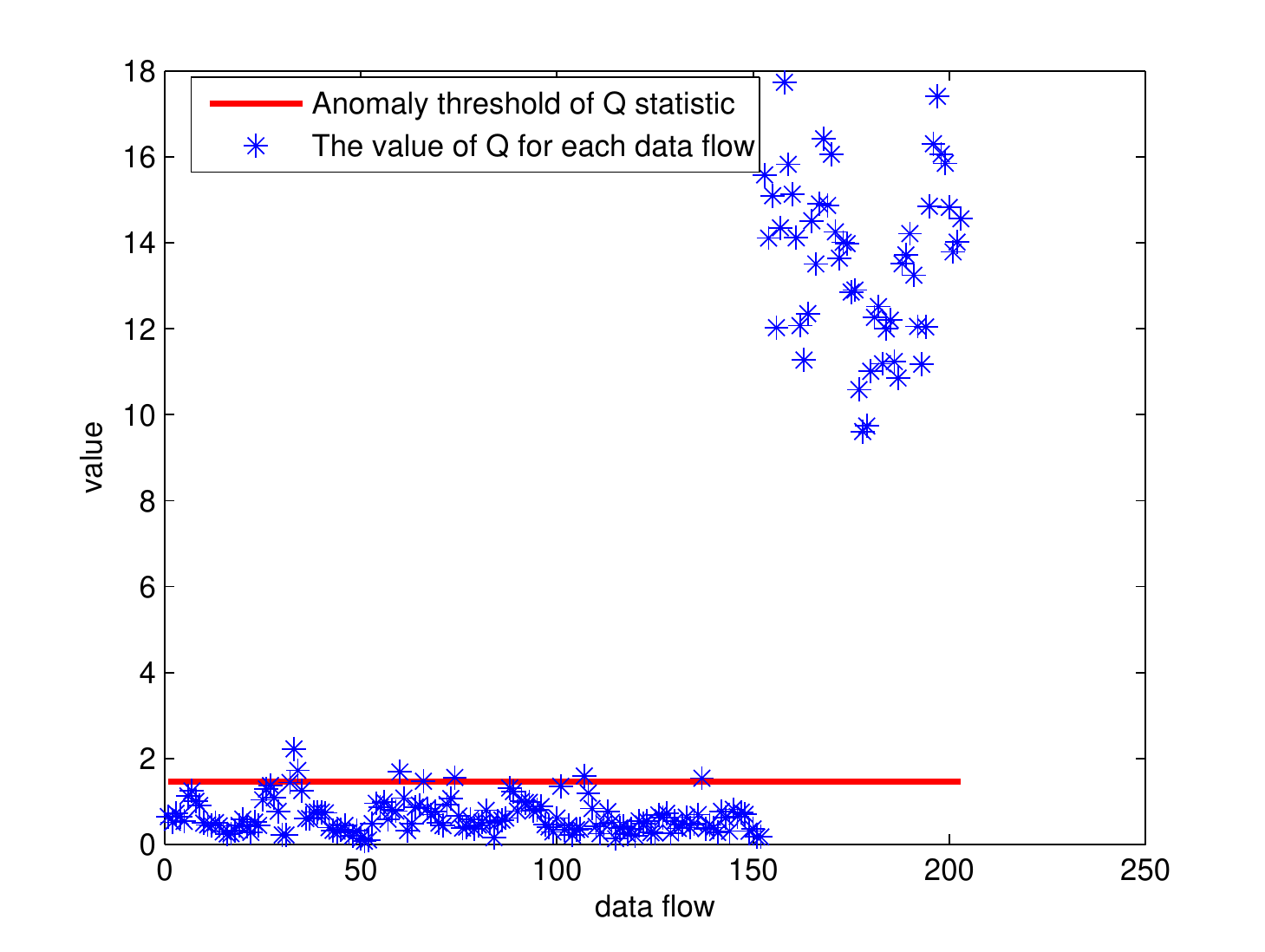}}

      \subfigure[The detection result of adding 1\% simulated errors based on $T^{2}$ statistic]{
      \label{fig:T1}
     \includegraphics[height=4.5cm,width=5cm]{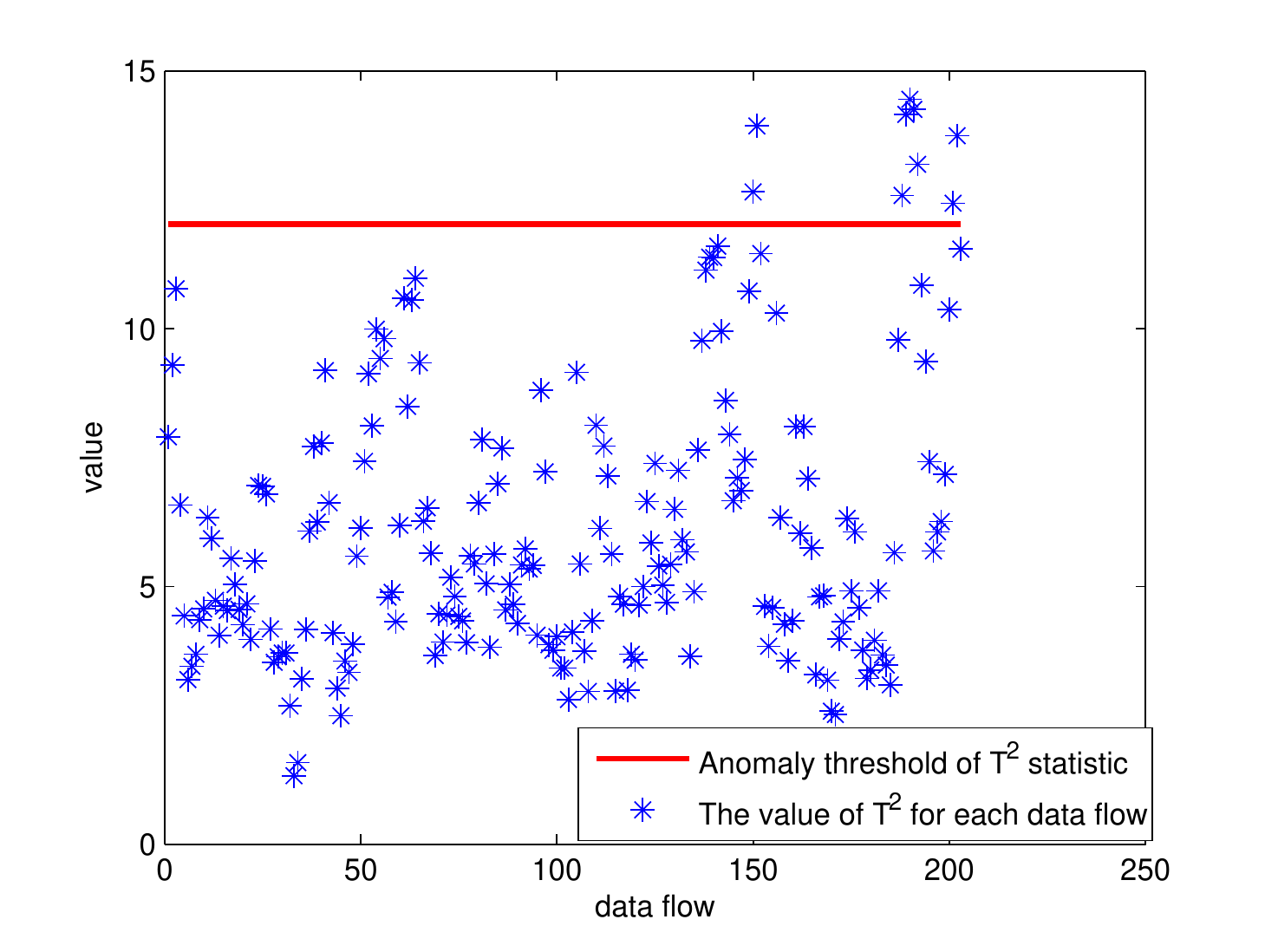}}
      \subfigure[The detection result of adding 5\% simulated errors based on $T^{2}$ statistic]{
      \label{fig:T2}
     \includegraphics[height=4.5cm,width=5cm]{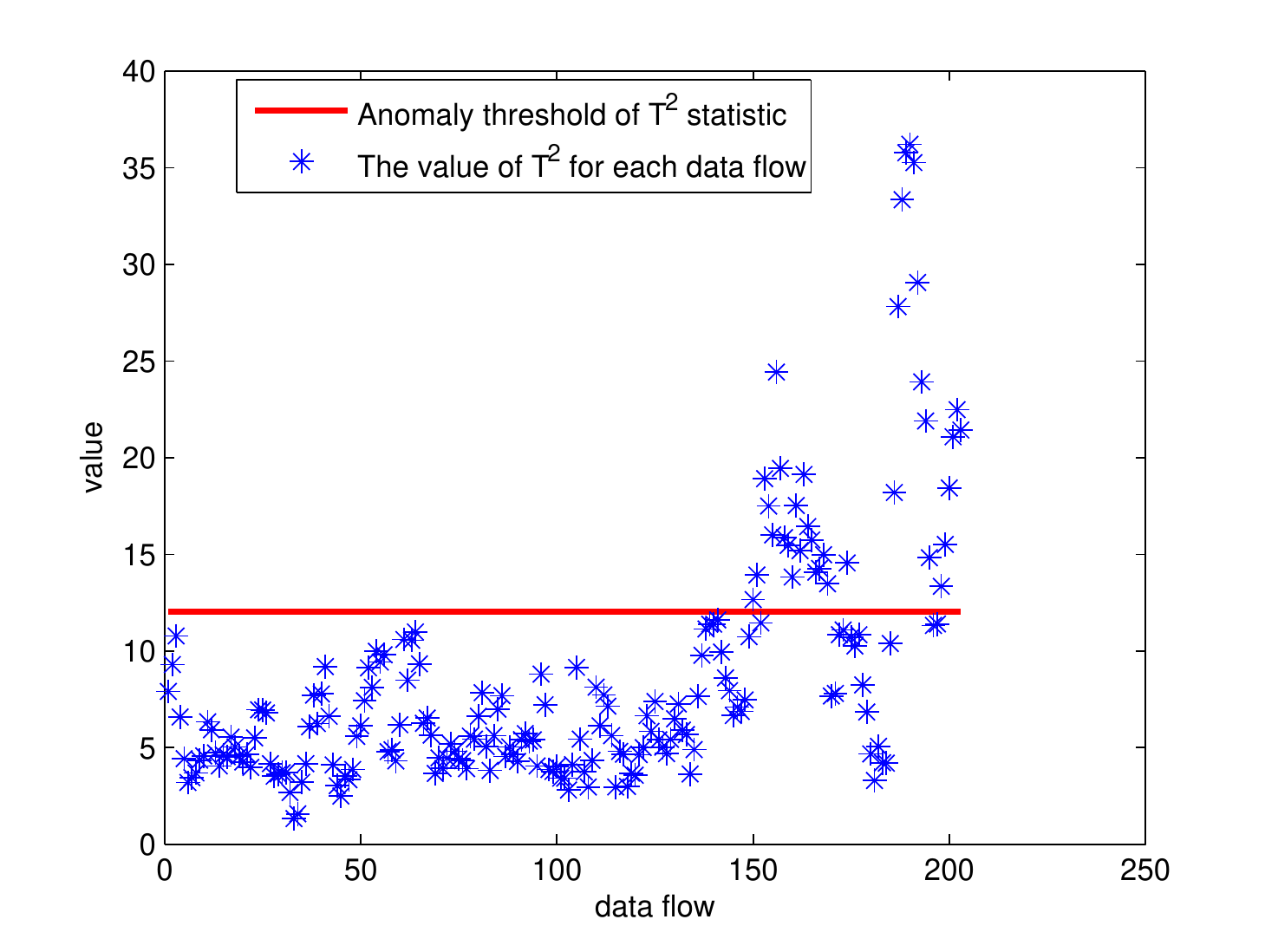}}
      \subfigure[The detection result of adding 10\% simulated errors based on $T^{2}$ statistic]{
      \label{fig:T3}
     \includegraphics[height=4.5cm,width=5cm]{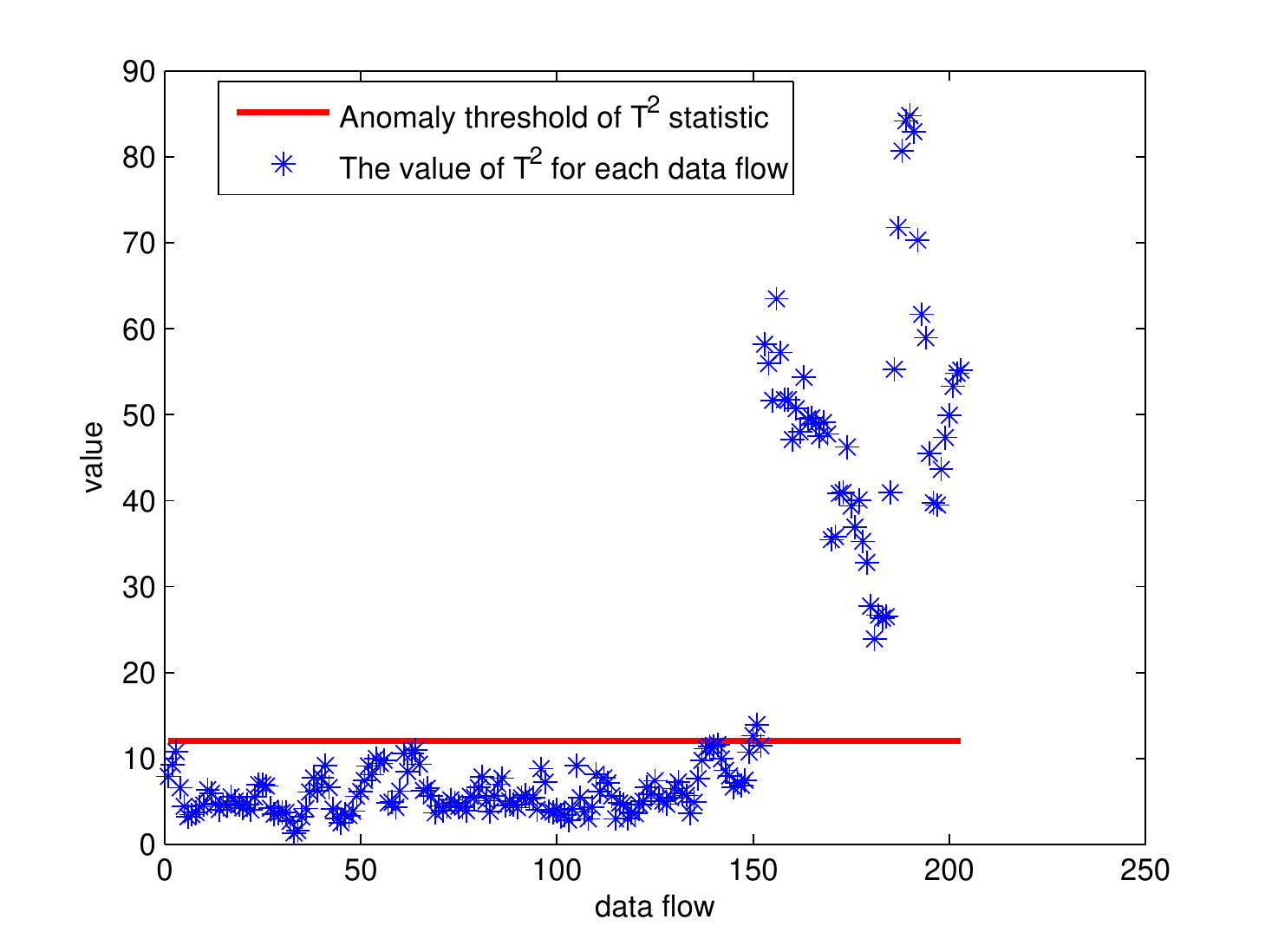}}
     \caption{The result of anomaly detection based on $Q$ and $T^{2}$ statistic methods.}
\end{figure*}

With the gathered sensor data, we can learn the structure of dependencies among those sensor nodes using the method described in Section 3. And Fig.\ref{fig:dependencies} shows the dependencies of 15 sensor nodes. In Fig.~\ref{fig:temp}, Fig.~\ref{fig:hum}, Fig.~\ref{fig:mic} horizontal axis denotes current node, and in vertical axis which is marked in stars denotes its parent nodes. Based on the learned structure of these sensor nodes the following experiments for sensor data anomaly and redundancy detection are conducted.
\subsection{The Result of Sensor Data Anomaly Detection}
For sensor data anomaly detection we have proposed an algorithm based on principal component analysis and Bayesian network. And this algorithm mainly contains two parts, in the first part, a rapid rough detection method based on principal statistic analysis is used to identify the data stream that may be abnormal. In the second part, with the method of Naive Bayesian classifier, we can find out the correct node that generates abnormal data.

In this simulation part, we first add artificial error data to the last 50 data streams in the collected 200 data streams. In order to generate abnormal data, on the basis of original data we use Eq.~\ref{eq14} to generate abnormal data.
\begin{equation}\label{eq14}
  X_{i,j}=X_{i,j}+ave_{j}\times P
\end{equation}
Where $X_{ij}$ denotes the original gathered sensor data of node $j$ at time $i$, $ave_{j}$ denotes the mean value of the training data for node $j$, $P$ denotes the error percentage.

Fig.~\ref{fig:Q1}-Fig.~\ref{fig:T3} shows the anomaly detection result of adding 1\%,5\%, and 10\% simulated errors based on $Q$ and $T^{2}$ statistic method. We can learn that the method of $T^{2}$ does not perform well when the added error is from 1\% to 5\%. And when the added error is 10\%, all the abnormal data streams are detected. It can be seen that the $Q$ statistic method is more sensitive than the $T^{2}$ statistic method, and it can detect all the abnormal data points when the added error is 5\%.

We have mentioned before that the anomaly detection method based on principal statistic analysis is a ``rough" detection. Through the above analysis of the results of the anomaly detection based on the two statistics, it is possible to improve the effectiveness of the ``rough" detection if the two statistics are combined and in an ``OR" relationship. And we name the ``rough" method as $TQ$ algorithm.

In order to measure the effectiveness and feasibility of the algorithm, we use ``precision" and ``recall" as the verification indicators.
\begin{equation}\label{eq15}
  precision=\frac{TP}{TP+FP}\quad recall=\frac{TP}{TP+FN}
\end{equation}
Where $TP$ denotes the number of positive cases that are judged as positive, $TN$ denotes the number of negative cases that are determined as negative, $FP$ denotes the number of negative cases that are judged as positive, $FN$ denotes the number of positive cases that are judged as negative.
\begin{figure}
\begin{minipage}[t]{0.5\linewidth}
\centering 
\includegraphics[height=5cm,width=6cm]{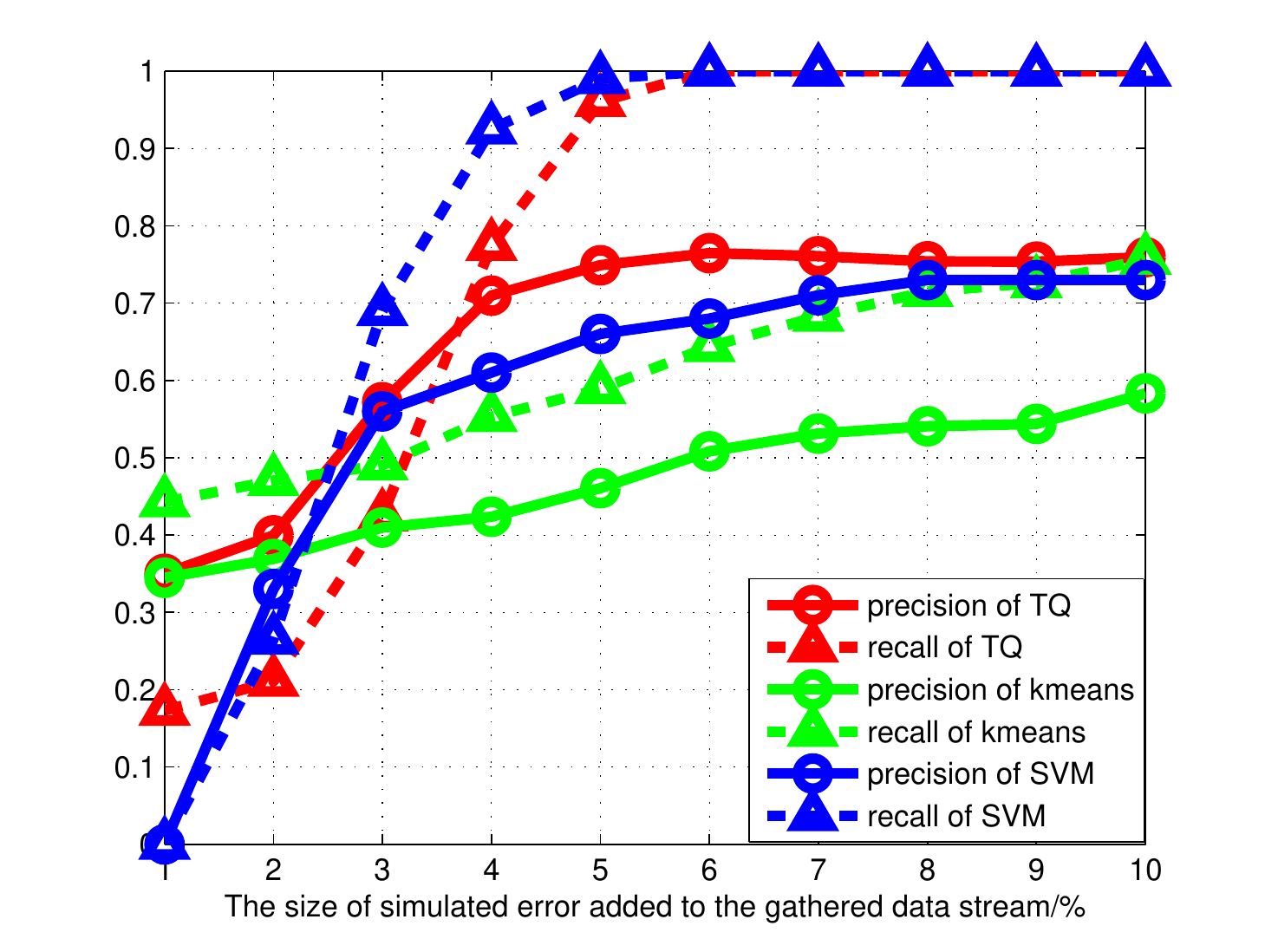}
\caption{The result of sensor data anomaly detection based on TQ method} 
\label{fig:TQ_Kmeans_svm}
\end{minipage}
\begin{minipage}[t]{0.5\linewidth}
\centering 
\includegraphics[height=5cm,width=6cm]{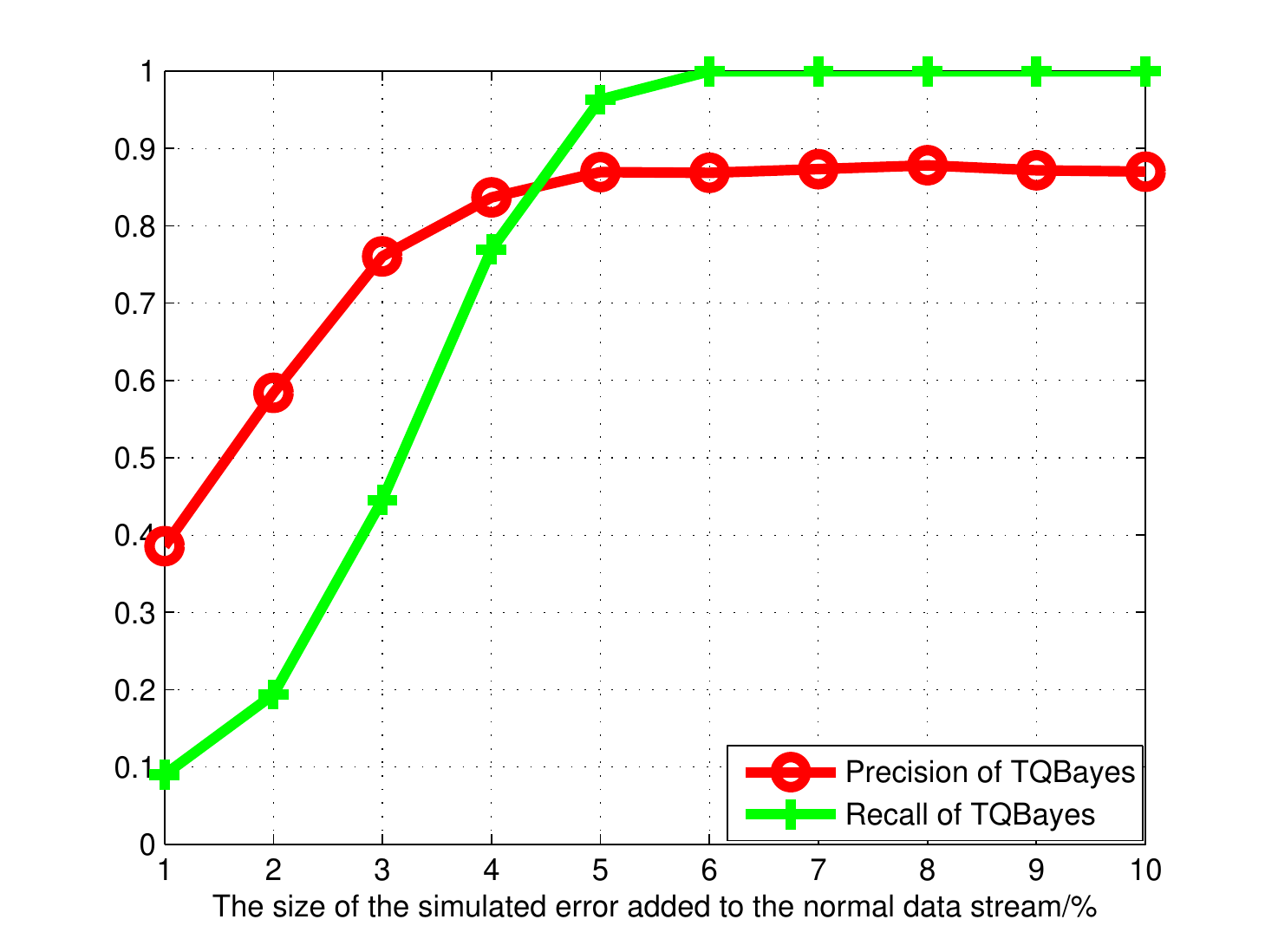}
\caption{The result of sensor data anomaly detection based on TQBayes}
\label{fig:TQ_Bayes}
\end{minipage}
\end{figure}
With the characteristics of the test datasets, the $TQ$ method can detect the anomaly of the data flow collected by 15 sensor nodes at each time in the ``rough" detection stage. The traditional anomaly detection algorithms such as methods based on SVM and K-means are commonly used in sensor data stream anomaly detection. Fig.~\ref{fig:TQ_Kmeans_svm} shows the comparison between the $TQ$ and other anomaly detection algorithms in the stage of rough detection. It can be seen that when the added simulation error is less than 6\%, from the perspective of the ``recall", the SVM-based approach is slightly better than the $TQ$ method. But from the point of view of ``precision", $TQ$ method is better than SVM-based method at all experimental points. Therefore, consider both ``recall" and ``precision", we can see that the $TQ$ method proposed in the ``rough" detection stage is feasible and effective.

In order to determine the anomaly data points, first of all, with the method in ``rough" detection stage the suspicious data flow can be determined, then based on the method proposed in ``careful" stage and the abnormal data points can be further identified. Fig.~\ref{fig:TQ_Bayes} shows the results of the $TQBayes$ based anomaly detection, it can be seen that the $TQ Bayes$ method improves the detection ``precision" in the case of ensuring the ``recall" compared with the stage of rough detection.
\subsection{The Result of Sensor Data Redundancy Elimination}
\begin{figure}[!t]
\centering
\subfigure[The redundant nodes of 15 temperature sensor nodes]{
\includegraphics[height=4cm,width=6cm,scale=0.5]{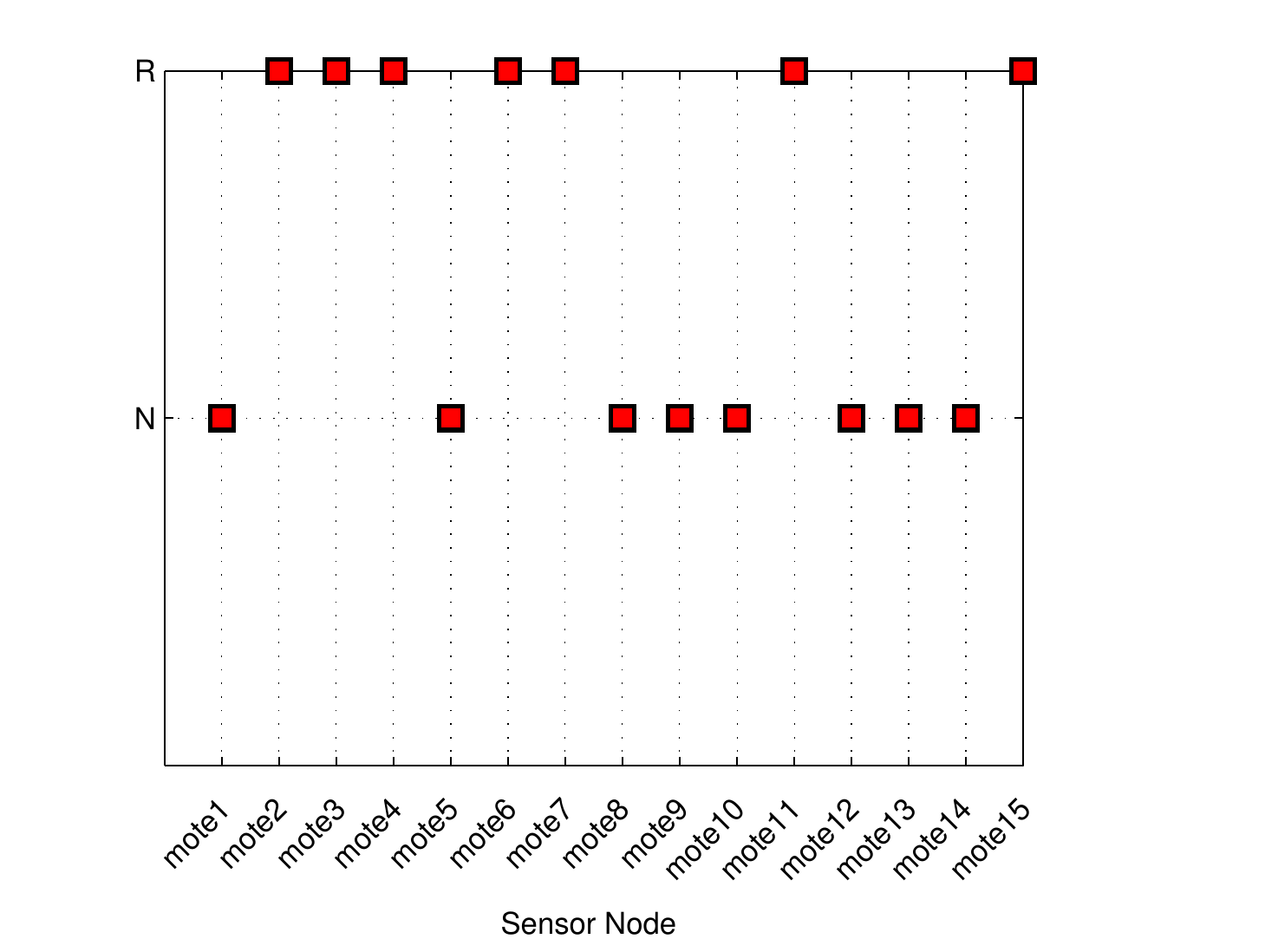}}
\subfigure[The redundant nodes of 15 humidity sensor nodes]{
\includegraphics[height=4cm,width=6cm,scale=0.5]{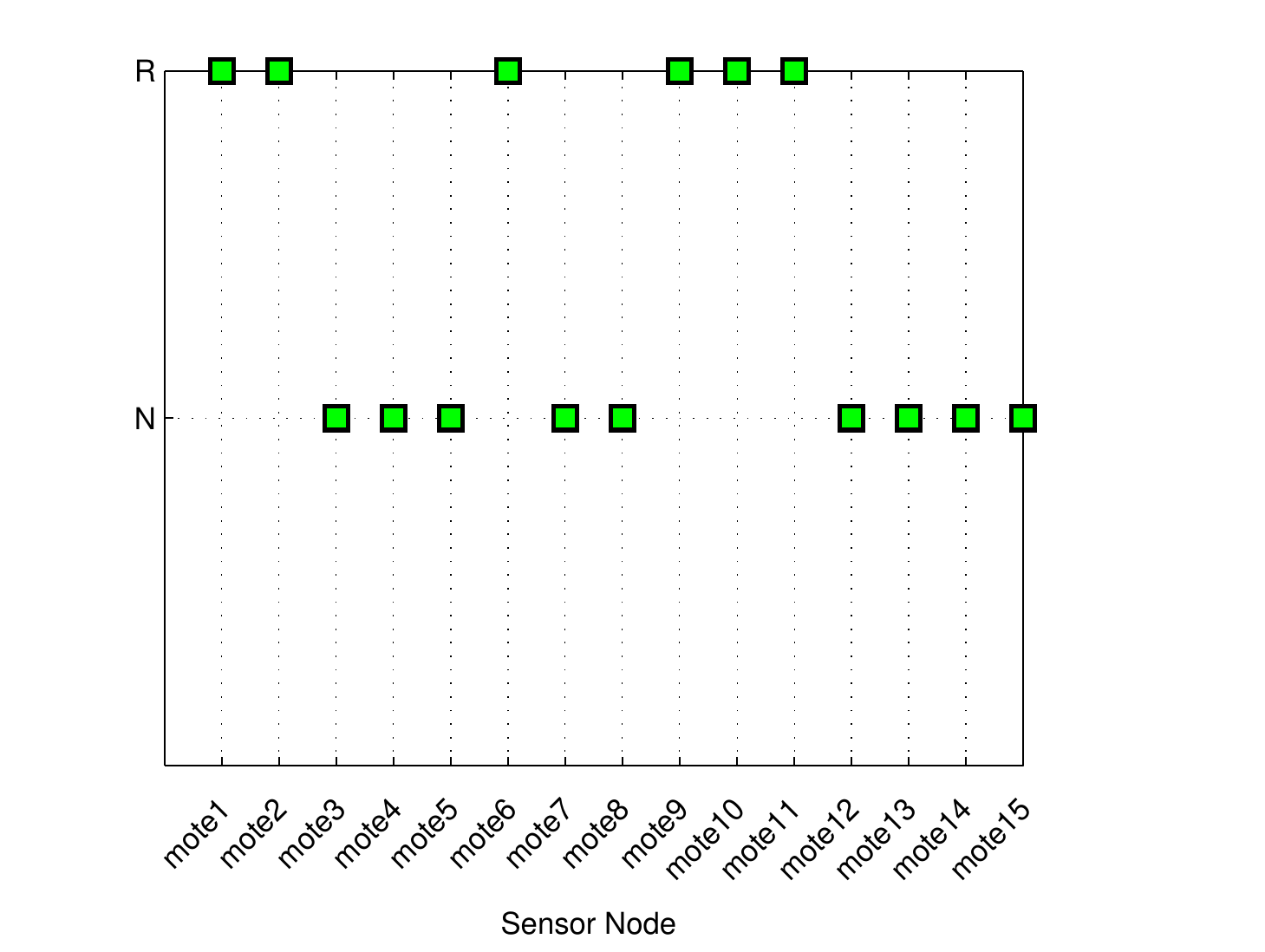}}
\subfigure[The redundant nodes of 15 microphone sensor nodes]{
\includegraphics[height=4cm,width=6cm,scale=0.5]{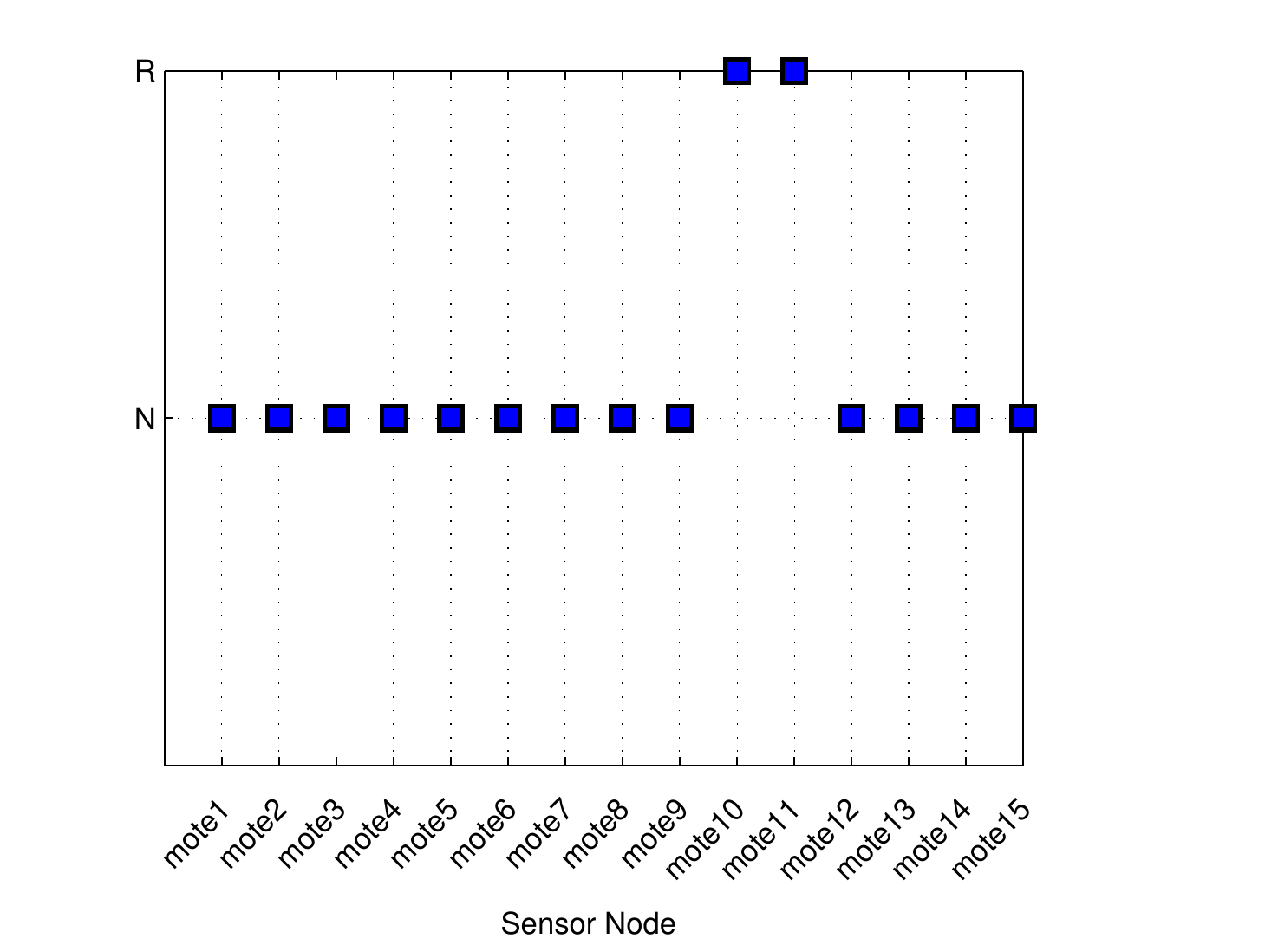}}
\subfigure[The number of redundant nodes with different number of sensor nodes]{
\includegraphics[height=4cm,width=6cm,scale=0.5]{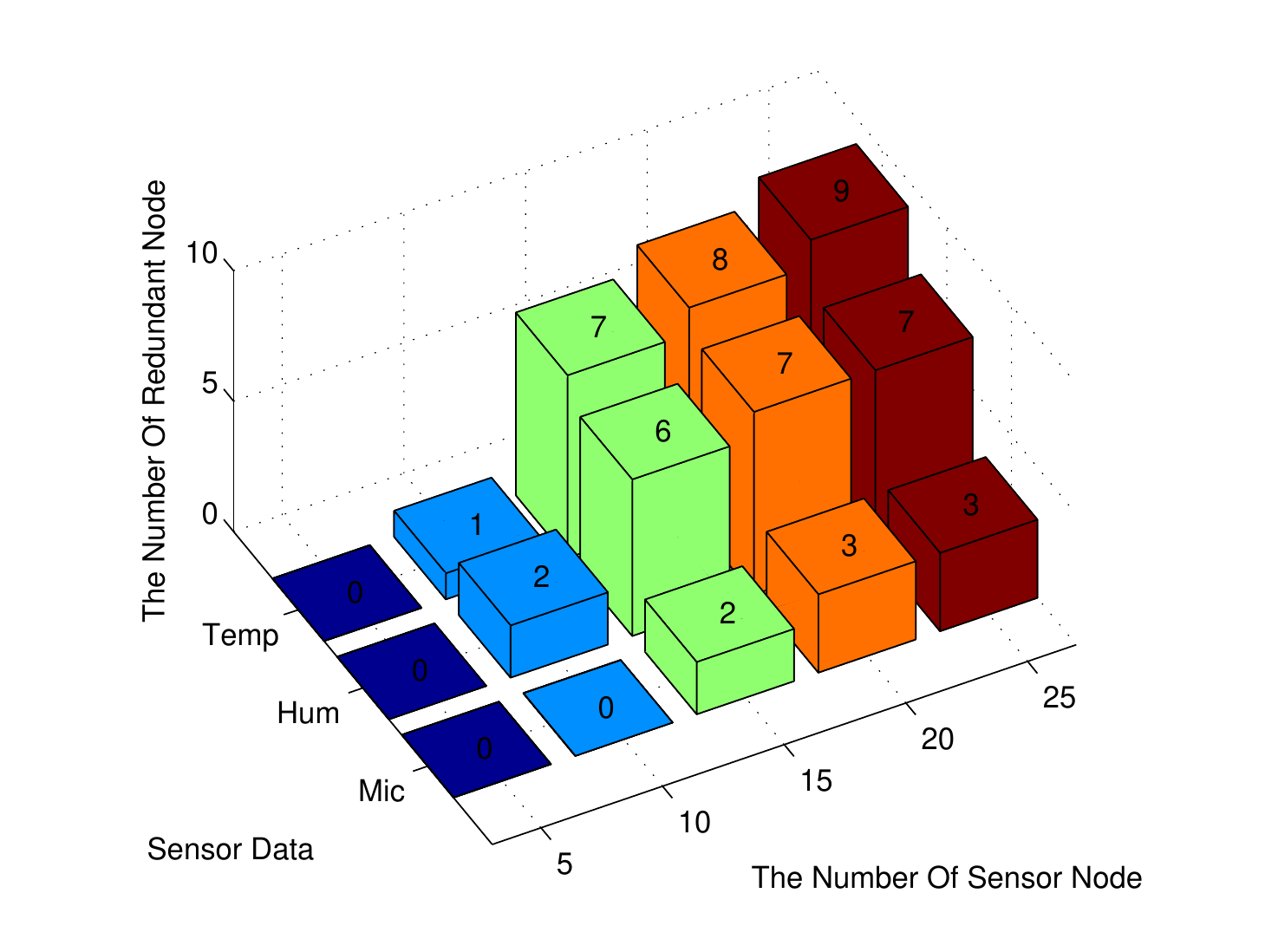}
\label{fig:number of redundant}
}
\caption{The result of redundant node detection based on SSDRDA (R denotes Redundant and N denotes non-redundant)}
\label{fig:result of resundant}
\end{figure}

In a gathered static dataset, if a specific node is detected as redundant node, it denotes the data collected by this node are redundant and we can get these data by its parent nodes. Based on this mechanism, we can weigh the performance of our algorithm by the accurately recovering the redundant data. The root-mean-square-error (RMSE) between real and predict values of redundant data is regarded as metrics.
\begin{equation}RMSE=\sqrt{mean[(y_{it}-\overline{y_{it}})^{2}]}\quad mean(RMSE)=\frac{\sum_{n=1}^{N}RMSE_{n}}{N}\end{equation}
Where $y_{it}$ is the real value of node $i$ at time $t$, $\overline{y_{it}}$ is the predicted value of node $i$ at time $t$, mean(RMSE) is the mean RMSE of all redundant nodes, $N$ is the number of redundant nodes.
\begin{figure}[!t]
\centering
\subfigure[The real and predicted temperature values from one redundant node]{
\includegraphics[height=4cm,width=8cm,scale=0.5]{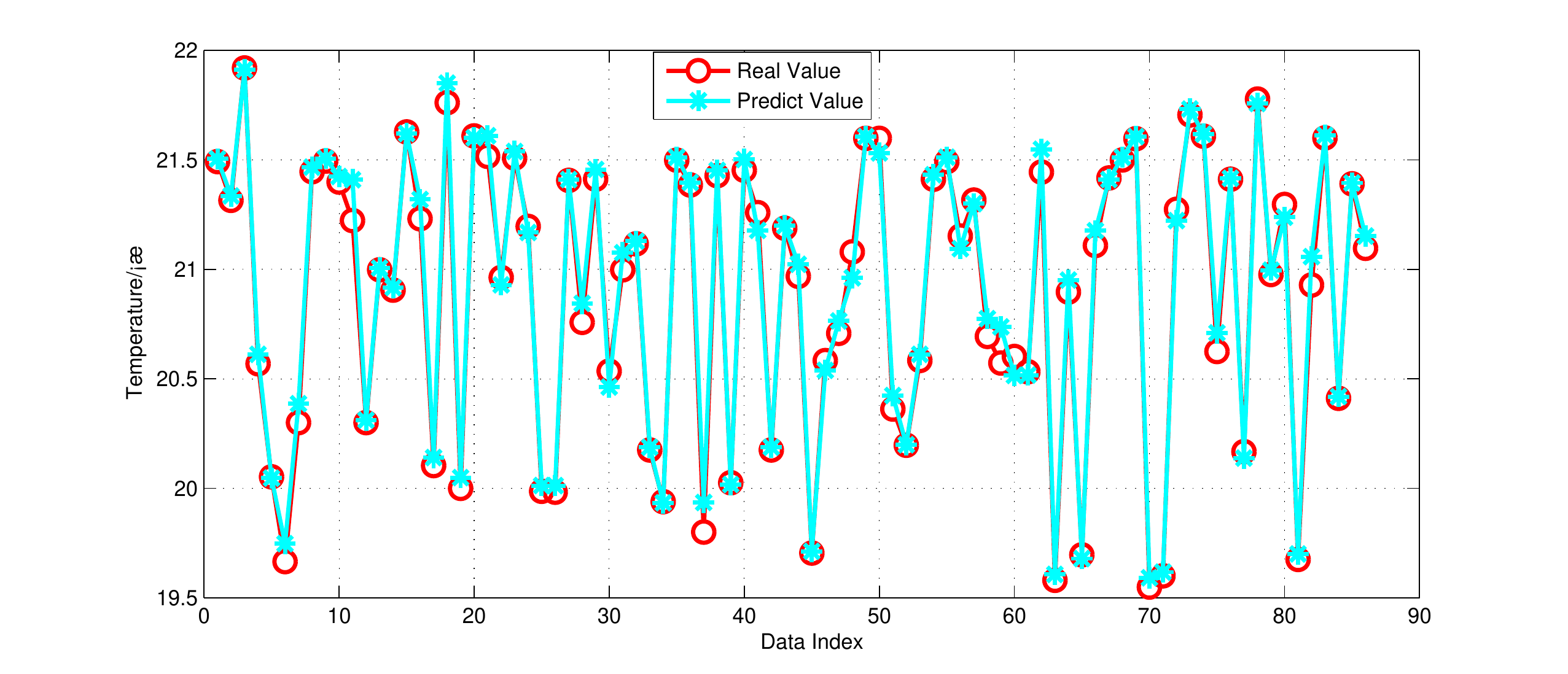}}
\subfigure[The real and predicted humidity values from one redundant node]{
\includegraphics[height=4cm,width=8cm,scale=0.5]{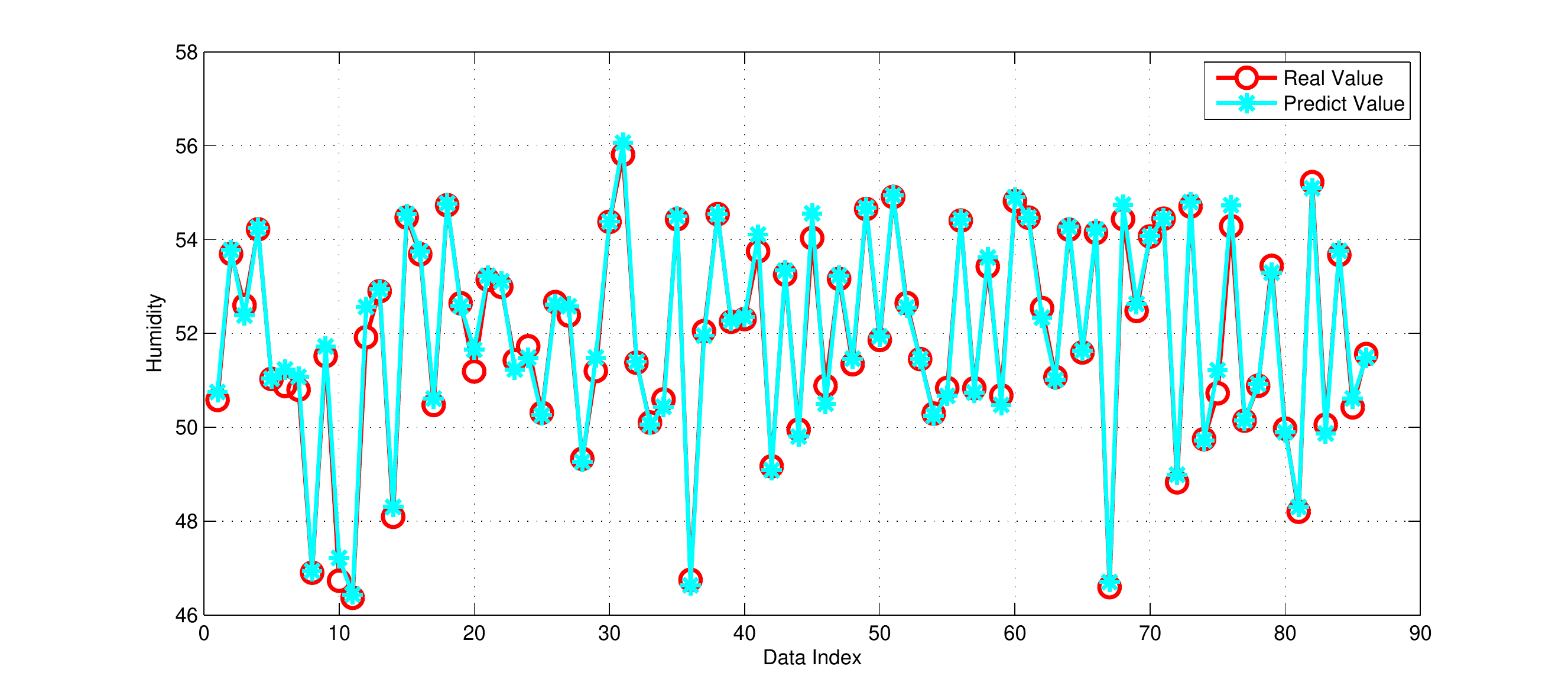}}
\subfigure[The real and predicted microphone values from one redundant node]{
\includegraphics[height=4cm,width=8cm,scale=0.5]{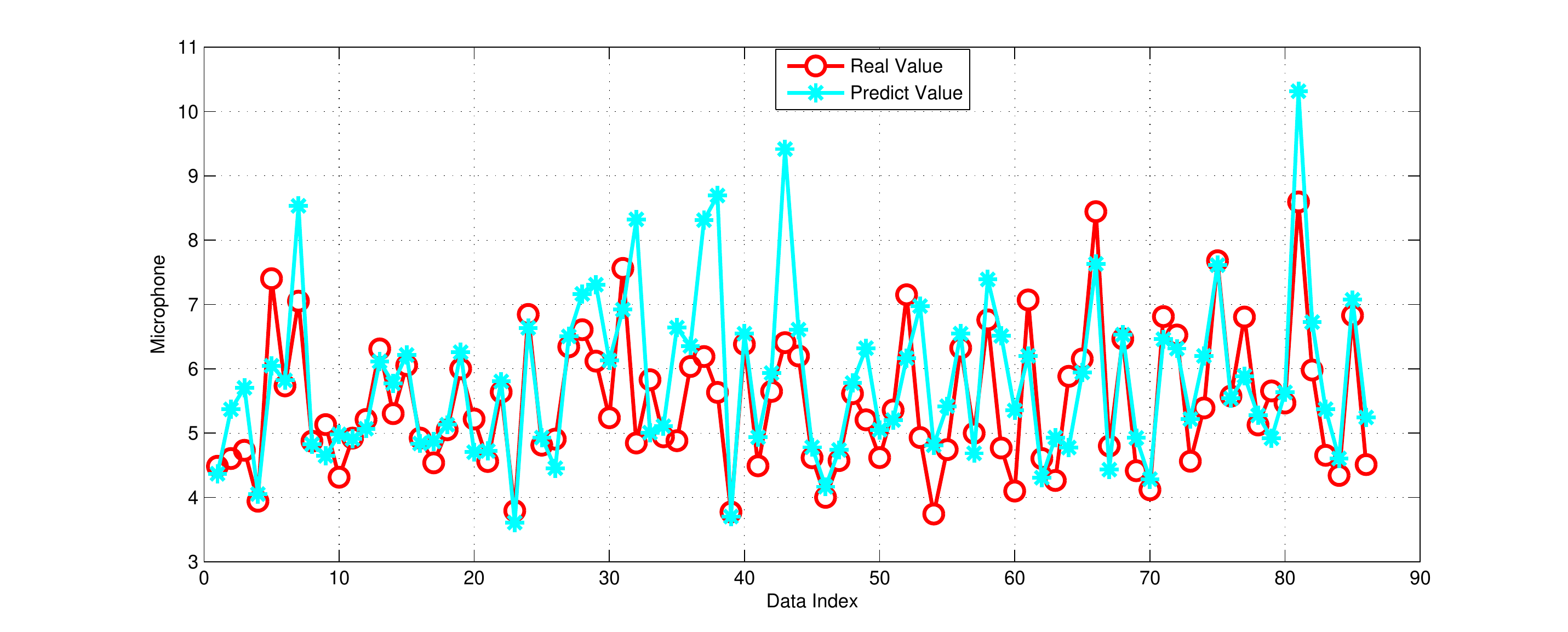}}
\subfigure[The mean RMSE value of all redundant nodes in SSDRDA]{
\label{fig:rmse}
\includegraphics[height=4cm,width=8cm,scale=0.5]{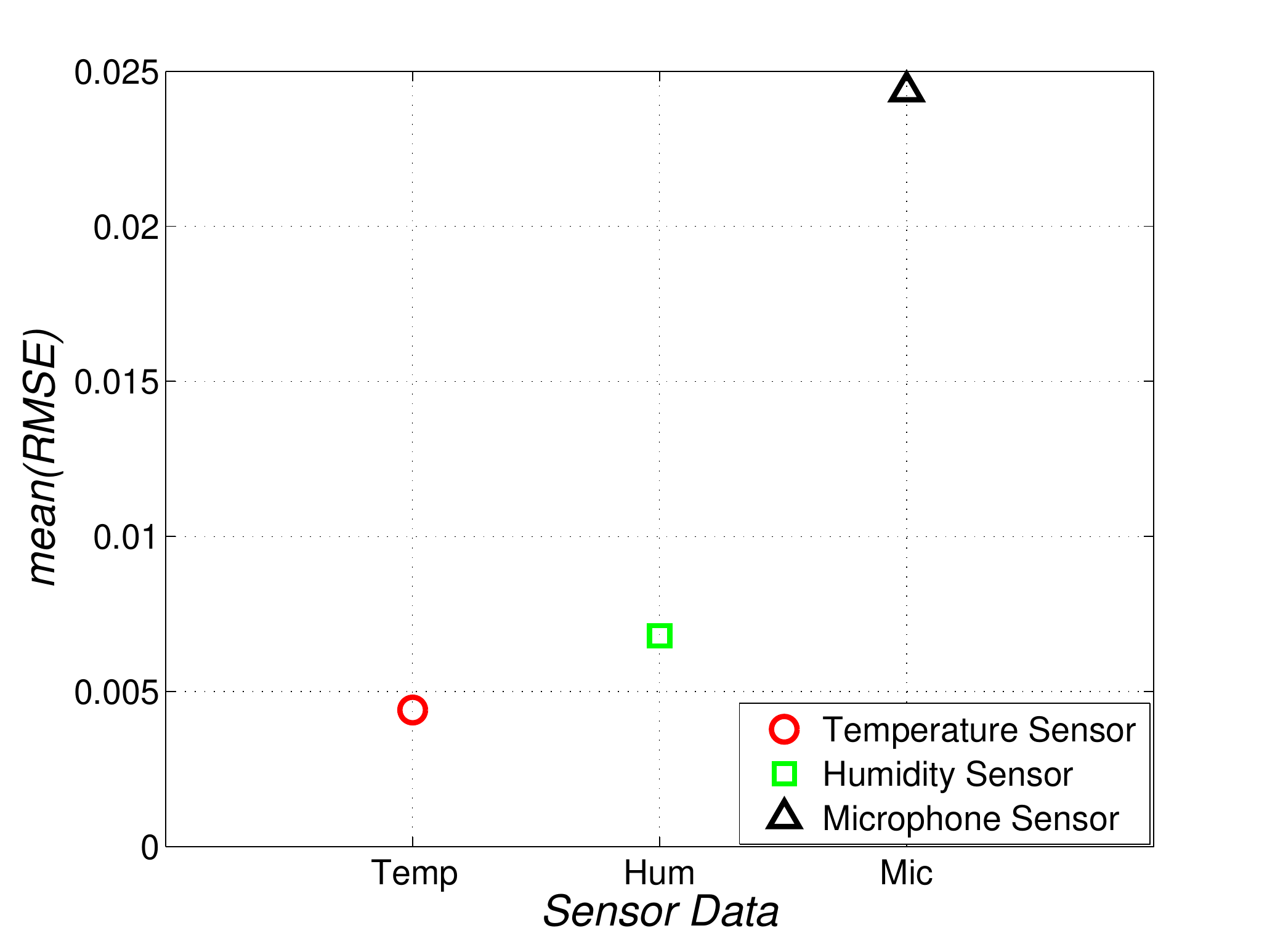}
}
\caption{The result of predicting the data of redundant node in a static data set}
\label{fig:predict result}
\end{figure}
\begin{figure}[!t]
\centering 
\includegraphics[height=4cm,width=10cm,scale=0.5]{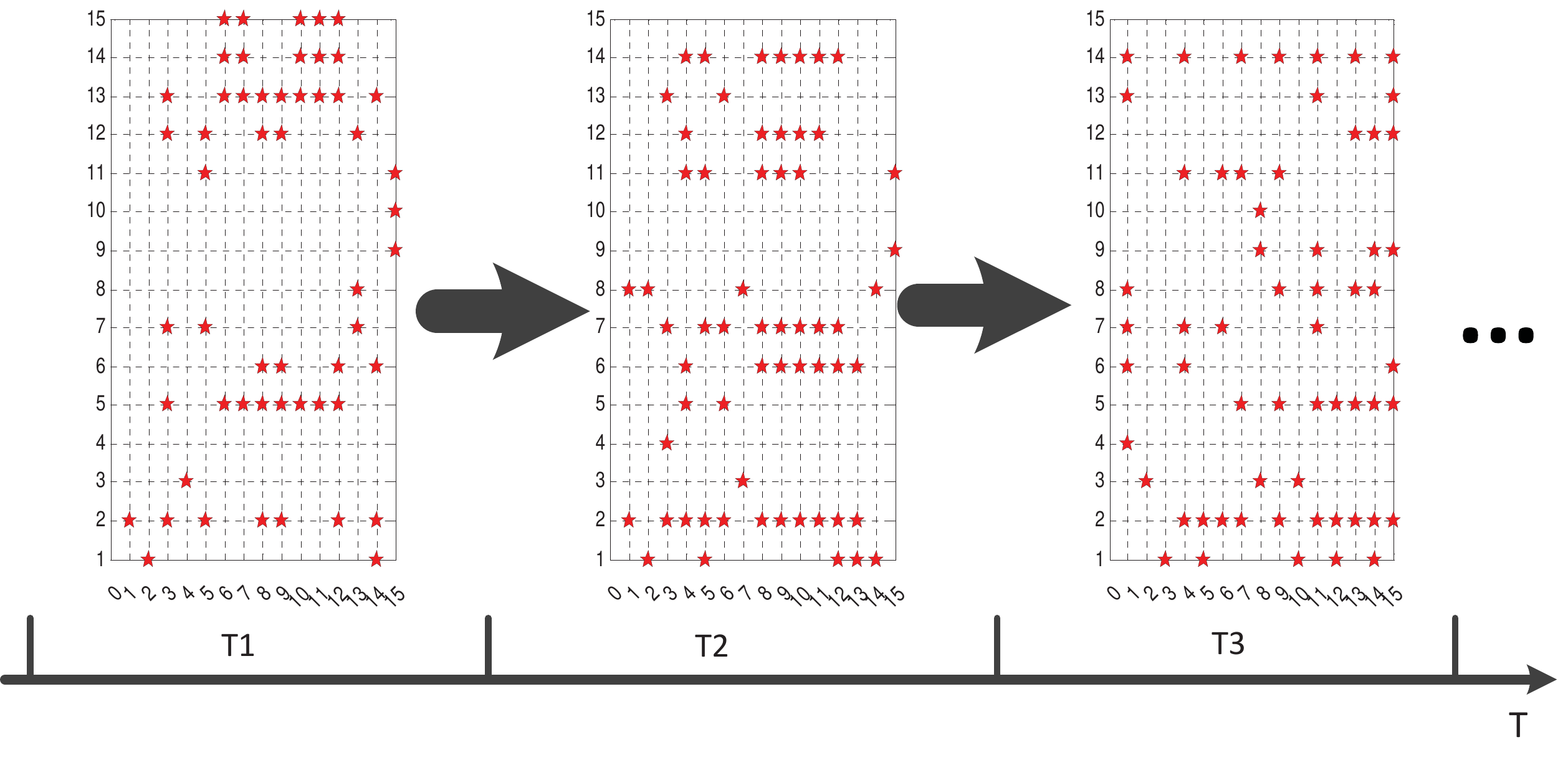}
\caption{The transition network in each time slot} 
\label{fig:transition}
\end{figure}
\begin{figure}[!t]
\centering
\subfigure[The predicted working state for parts of the temperature sensor nodes]{
\label{fig:temp1}
\includegraphics[height=5cm,width=7cm,scale=0.5]{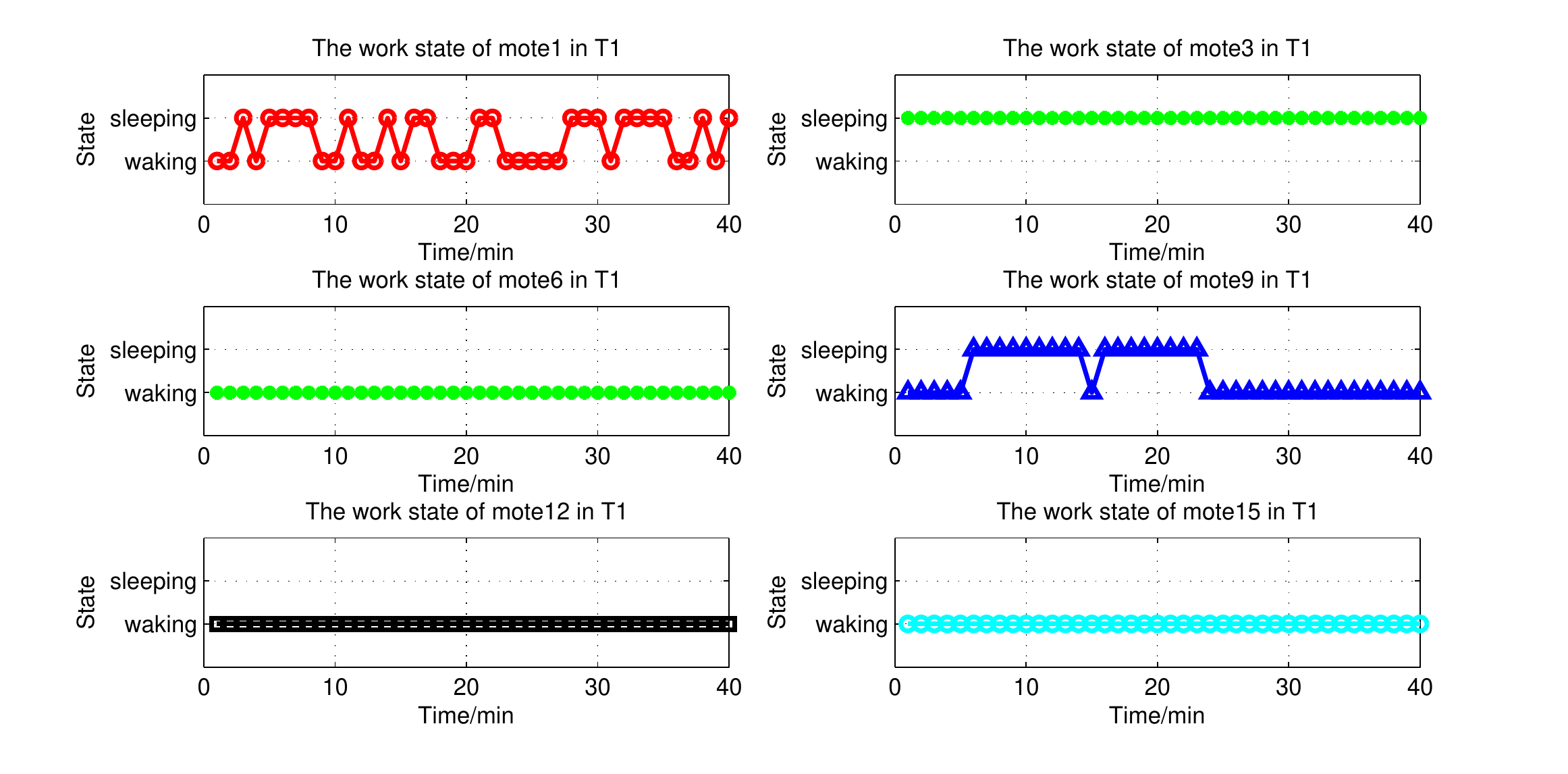}
}
\subfigure[The predicted working state for parts of the humidity sensor nodes]{
\label{fig:hum1}
\includegraphics[height=5cm,width=7cm,scale=0.5]{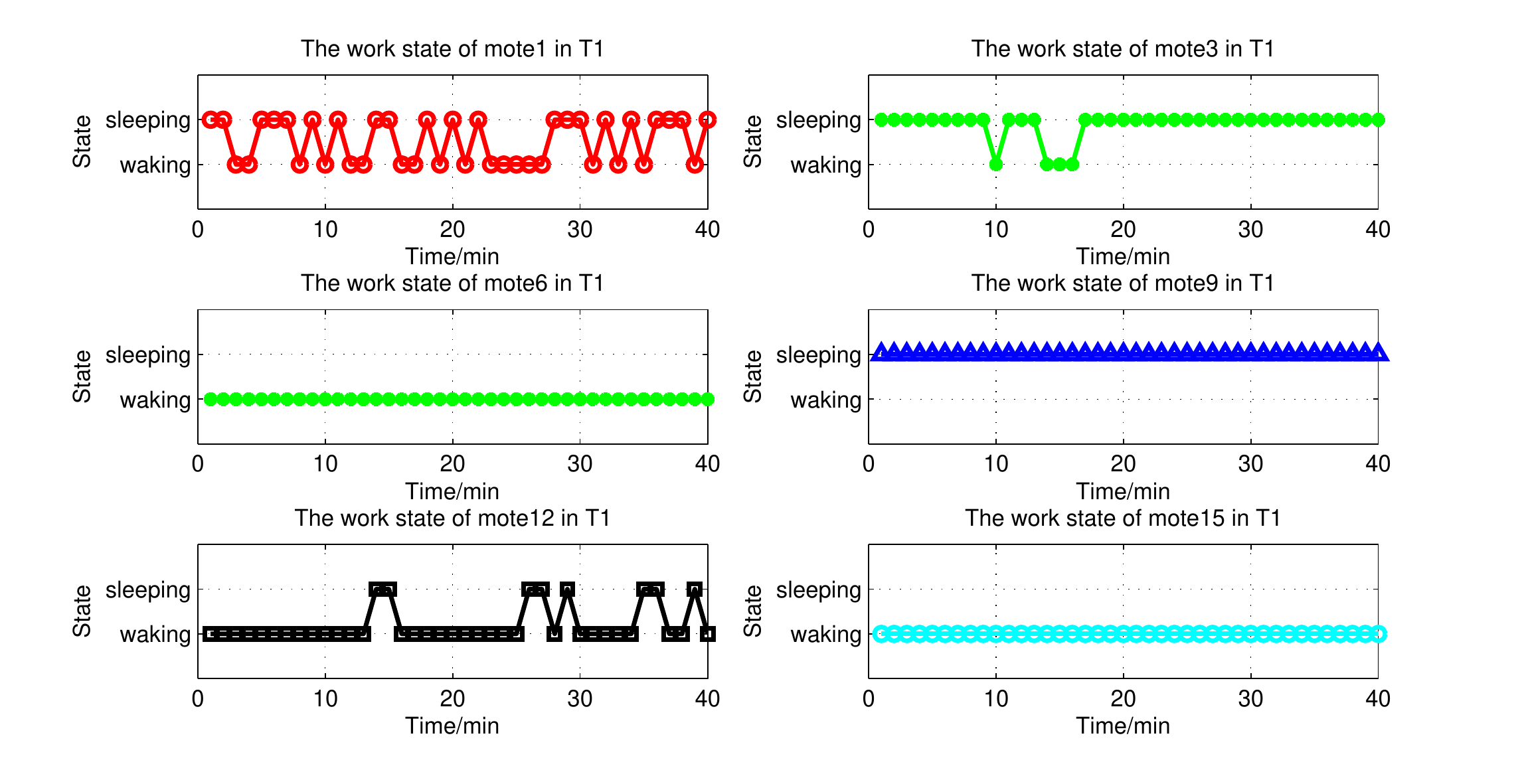}
}
\subfigure[The predicted working state for parts of the microphone sensor nodes]{
\label{fig:mic1}
\includegraphics[height=5cm,width=7cm,scale=0.5]{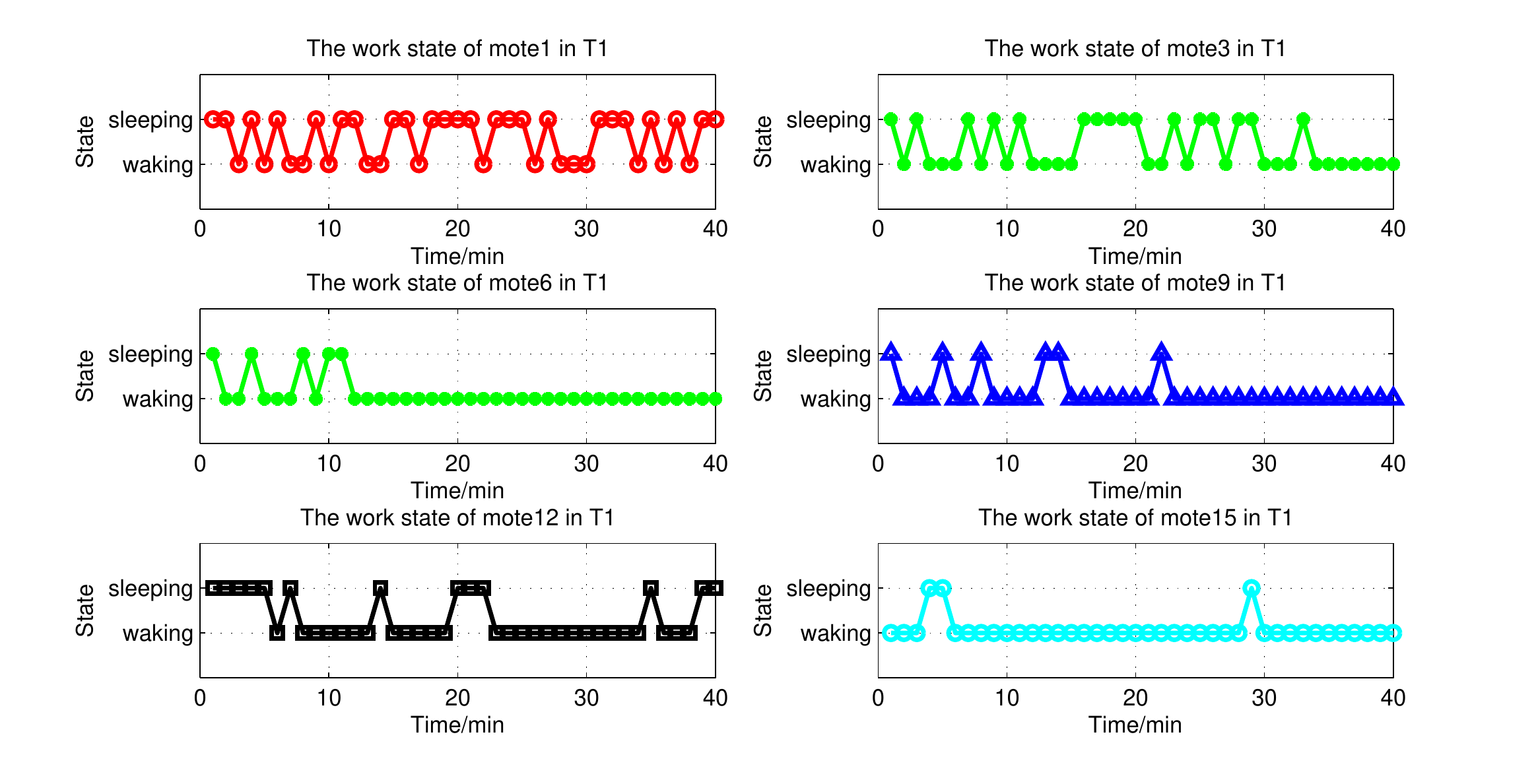}
}
\subfigure[The mean RMSE value of all redundant nodes based on RSDRDA]{
\label{fig:the RMSE of RSDRDA}
\includegraphics[height=4cm,width=7cm,scale=0.5]{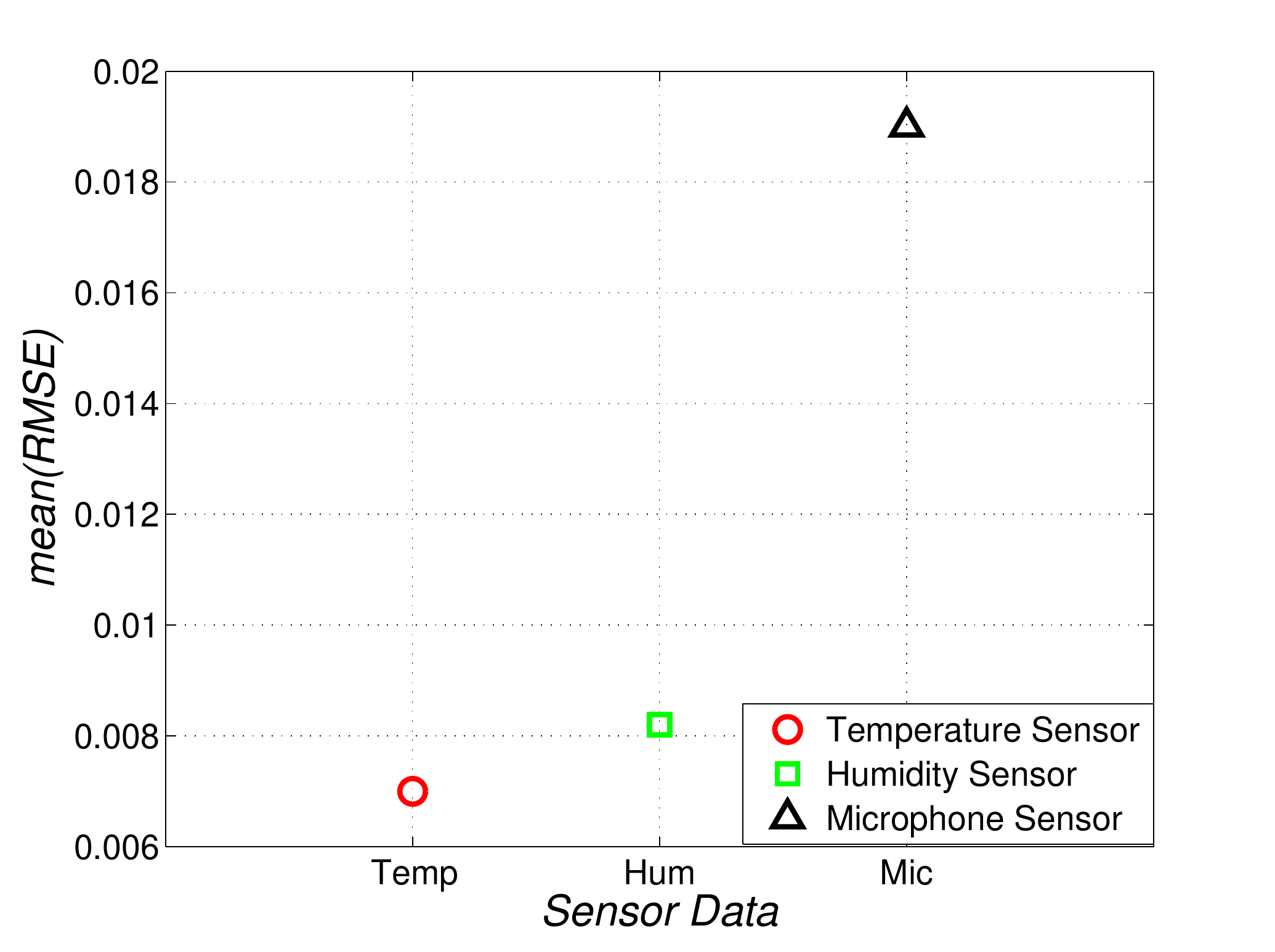}
}
\caption{The results of predicted state for parts of the sensor nodes based on RSDRDA}
\end{figure}

By applying static sensor data to the proposed algorithm, we can get the result of redundant node detection in the datasets of temperature, humidity, and microphone as shown in Fig.~\ref{fig:result of resundant}. And Fig.~\ref{fig:number of redundant} shows the number of redundant nodes in different total number of sensor nodes. We can learn that the number of redundant nodes in temperature and humidity data set is more than in microphone. The reason is that the gathered temperature and humidity data is gradually changed, and sensor nodes at different position may collect similar data. But the data of microphone sensor nodes collect is closely relevant to the position of the nodes and its data fluctuation is higher than temperature and humidity. Thus, the number of redundant nodes in temperature and humidity is more than in microphone.

In our datasets there is no prior knowledge to clearly divide the set into redundant and non-redundant parts, so the effectiveness of the detected redundant sensor nodes is hard to be conducted in terms of recall and precision. In order to get the effectiveness of the algorithm, we estimate the data which is detected as redundant and the RMSE of estimating result can reflect the accuracy of our algorithm. Considering the purpose of estimating redundant data is just to validate the feasible of SSDRDA, we use the most common and simple method which named weight method for missing data estimation. In this method we put different weight on the parent of redundant node as shown in Eq.~(\ref{eq:weight}), and the weight is base on the similarity between current node and its parent.
\begin{equation}\label{eq:weight}R=\frac{1}{W}[\sum_{k=1}^{N}(\frac{1}{d_{k}}X_{k})]\end{equation}
where $\frac{1}{d_{k}}$ is the weight, $W=\sum_{k=1}^{N}\frac{1}{d_{k}}$.

Fig.~\ref{fig:predict result} shows the result of predicting the data of redundant node by its parent nodes. And Fig.~\ref{fig:rmse} shows the mean RMSE of predicting and real value. Because of the fluctuation of the data collected by microphone mote is higher than that of temperature and humidity, and we can learn that the error rate of microphone data is higher than that of temperature and humidity.

We have proposed RSDRDA for the real-time redundancy detection. Fig.~\ref{fig:transition} shows the process of variable transition network of fifteen temperature sensor nodes. We define 100 minutes as one time slice, and in the first 60 minutes all of sensor nodes are in ``waking" state, then use the data sets collected in first 60 minutes to build the transition network as shown in Fig.~\ref{fig:transition}. Fig.~\ref{fig:temp1}-\ref{fig:mic1} shows the predict working state of parts of the sensor nodes in the last 40 minutes. If the predicted state of one sensor node is ``sleeping" at a specific time, it means the node can be sleeping at this time, otherwise, it will generate redundant data. If the predicted state of one sensor node is ``waking", then the node needs to wake at this time. In order to validate the accuracy of the predicted state, we recover the redundant data. And Fig.~\ref{fig:the RMSE of RSDRDA} shows the mean RMSE of real and estimated data of all redundant data. From Fig.~\ref{fig:the RMSE of RSDRDA} we can learn that the RSDRDA is good at real-time redundancy detection.

\section{Conclusion}
This paper investigates the preprocessing methods for big sensor data in IoT. And a framework that is composed of two parts for sensor data anomaly and redundancy detection has been proposed. In the first part, an algorithm based on principal statistic analysis and Bayesian networks has been proposed for sensor data anomaly detection. From the result of comparing with other traditional anomaly detection algorithms it can be seen that our algorithm can improve the precision of anomaly detection while ensuring the result of recall. In the second part, approaches based on SBN and DBNs are proposed for sensor data elimination. We have proposed SSDRDA to eliminate redundant data in a static data set and RSDRDA to eliminate redundant sensor data in real-time. And the RSDRDA is based on a new time-varying DBN model that is capable of describing the evolution of nonstationary temporal sequences. In order to validate the accuracy of proposed algorithm, we use a common method to recover redundant data, and the result of RMSE shows that the proposed algorithms are feasible and effective.
\section*{Acknowledgment}
This work is supported by the Fundamental Research Funds for the Central Universities (N140404015).
\bibliographystyle{ieeetr}
\bibliography{myreference}

\begin{thebibliography}{10}

\bibitem{Tsai2014Data}
C.~W. Tsai, C.~F. Lai, M.~C. Chiang, and L.~T. Yang, ``Data mining for internet
  of things: A survey,'' {\em IEEE Communications Surveys \& Tutorials},
  vol.~16, no.~1, pp.~77--97, 2014.

\bibitem{Tsai2014Future}
C.~W. Tsai, C.~F. Lai, and A.~V. Vasilakos, ``Future internet of things: open
  issues and challenges,'' {\em Wireless Networks}, vol.~20, no.~8,
  pp.~2201--2217, 2014.

\bibitem{Chen2014Big}
M.~Chen, S.~Mao, and Y.~Liu, ``Big data: A survey,'' {\em Mobile Networks and
  Applications}, vol.~19, no.~2, pp.~171--209, 2014.

\bibitem{fateh2013energy}
B.~Fateh and M.~Govindarasu, ``Energy minimization by exploiting data
  redundancy in real-time wireless sensor networks,'' {\em Ad Hoc Networks},
  vol.~11, no.~6, pp.~1715--1731, 2013.

\bibitem{marti2015anomaly}
L.~Mart{\'\i}, N.~Sanchez-Pi, J.~M. Molina, and A.~C.~B. Garcia, ``Anomaly
  detection based on sensor data in petroleum industry applications,'' {\em
  Sensors}, vol.~15, no.~2, pp.~2774--2797, 2015.

\bibitem{illiano2015detecting}
V.~P. Illiano and E.~C. Lupu, ``Detecting malicious data injections in wireless
  sensor networks: A survey,'' {\em ACM Computing Surveys (CSUR)}, vol.~48,
  no.~2, p.~24, 2015.

\bibitem{chen2015distributed}
P.-Y. Chen, S.~Yang, and J.~A. McCann, ``Distributed real-time anomaly
  detection in networked industrial sensing systems,'' {\em IEEE Transactions
  on Industrial Electronics}, vol.~62, no.~6, pp.~3832--3842, 2015.

\bibitem{Salem2014Anomaly}
O.~Salem, A.~Guerassimov, A.~Mehaoua, A.~Marcus, and B.~Furht, ``Anomaly
  detection in medical wireless sensor networks using svm and linear regression
  models,'' {\em International Journal of E-Health and Medical Communications},
  vol.~5, no.~1, pp.~20--45, 2014.

\bibitem{martins2015support}
H.~Martins, L.~Palma, A.~Cardoso, and P.~Gil, ``A support vector machine based
  technique for online detection of outliers in transient time series,'' in
  {\em Proceedings of IEEE 10th Asian Control Conference (ASCC)}, pp.~1--6,
  2015.

\bibitem{Xiao2008Attribute}
H.~Xiao and Wang, ``Attribute selection-based and support vector machine for
  anomaly detection,'' {\em Journal of Huazhong University of Science \&
  Technology}, vol.~36, no.~3, pp.~99--102, 2008.

\bibitem{Shaikh2014Efficient}
S.~A. Shaikh and H.~Kitagawa, ``Efficient distance-based outlier detection on
  uncertain datasets of gaussian distribution,'' {\em World Wide Web}, vol.~17,
  no.~4, pp.~511--538, 2014.

\bibitem{Zhao2014Adaptive}
J.~Zhao, K.~Liu, W.~Wang, and Y.~Liu, ``Adaptive fuzzy clustering based anomaly
  data detection in energy system of steel industry,'' {\em Information
  Sciences}, vol.~259, no.~3, pp.~335--345, 2014.

\bibitem{shaikh2014top}
S.~A. Shaikh and H.~Kitagawa, ``Top-k outlier detection from uncertain data,''
  {\em International Journal of Automation and Computing}, vol.~11, no.~2,
  pp.~128--142, 2014.

\bibitem{Ma2016Supervised}
J.~Ma, L.~Sun, H.~Wang, Y.~Zhang, and U.~Aickelin, ``Supervised anomaly
  detection in uncertain pseudoperiodic data streams,'' {\em ACM Transactions
  on Internet Technology}, vol.~16, no.~1, pp.~1--20, 2016.

\bibitem{Angiulli2009DOLPHIN}
F.~Angiulli and F.~Fassetti, ``Dolphin: An efficient algorithm for mining
  distance-based outliers in very large datasets,'' {\em ACM Transactions on
  Knowledge Discovery from Data (TKDD)}, vol.~3, no.~1, pp.~777--781, 2009.

\bibitem{Moshtaghi2011Clustering}
M.~Moshtaghi, T.~C. Havens, J.~C. Bezdek, L.~Park, C.~Leckie, S.~Rajasegarar,
  J.~M. Keller, and M.~Palaniswami, ``Clustering ellipses for anomaly
  detection,'' {\em Pattern Recognition}, vol.~44, no.~1, pp.~55--69, 2011.

\bibitem{Huang2014Physics}
H.~Huang, H.~Qin, S.~Yoo, and D.~Yu, ``Physics-based anomaly detection defined
  on manifold space,'' {\em ACM Transactions on Knowledge Discovery from Data},
  vol.~9, no.~2, pp.~1--39, 2014.

\bibitem{Liu2012Isolation}
F.~T. Liu, K.~M. Ting, and Z.~H. Zhou, ``Isolation-based anomaly detection,''
  {\em ACM Transactions on Knowledge Discovery from Data}, vol.~6, no.~1,
  pp.~74--77, 2012.

\bibitem{Sch2001Estimating}
B.~Schlkopf, J.~Platt, J.~Shawe-Taylor, and A.~Smola, ``Estimating the support
  of a high-dimensional distribution,'' {\em Neural Computation}, vol.~13,
  no.~7, pp.~1443--1471, 2001.

\bibitem{Liang2005Redundancy}
Q.~Liang and L.~Wang, ``Redundancy reduction in wireless sensor networks using
  svd-qr,'' in {\em Proceedings of IEEE Military Communications Conference},
  pp.~1857--1861 Vol. 3, 2005.

\bibitem{vinas2015redundancy}
G.~Vi{\~n}as~Raventos, ``Redundancy elimination for data aggregation in
  wireless sensor networks,'' {\em \url{http://hdl.handle.net/2117/79900}}.

\bibitem{Coudert2015Robust}
D.~Coudert, A.~Kodjo, and T.~K. Phan, ``Robust energy-aware routing with
  redundancy elimination,'' {\em Computers \& Operations Research}, vol.~64,
  no.~C, pp.~71--85, 2015.

\bibitem{Patil2010SVM}
P.~Patil and U.~Kulkarni, ``Svm based data redundancy elimination for data
  aggregation in wireless sensor networks,'' {\em Wireless Sensor Network},
  vol.~2, no.~4, pp.~300--308, 2010.

\bibitem{Khedo2010READA}
K.~Khedo, R.~Doomun, and S.~Aucharuz, ``Reada: Redundancy elimination for
  accurate data aggregation in wireless sensor networks,'' {\em Wireless Sensor
  Network}, vol.~2, no.~4, pp.~300--308, 2010.

\bibitem{Ghanmy2011Characterization}
N.~Ghanmy, M.~A. Mahjoub, and N.~E.~B. Amara, ``Characterization of dynamic
  bayesian network,'' {\em International Journal of Advanced Computer Science
  \& Applications}, vol.~2, no.~7, 2011.

\bibitem{Sun2015A}
J.~Sun and J.~Sun, ``A dynamic bayesian network model for real-time crash
  prediction using traffic speed conditions data,'' {\em Transportation
  Research Part C Emerging Technologies}, vol.~54, pp.~176--186, 2015.

\bibitem{larranaga2013a}
P.~Larranaga, H.~Karshenas, C.~Bielza, and R.~Santana, ``A review on
  evolutionary algorithms in bayesian network learning and inference tasks,''
  {\em Information Sciences}, vol.~233, pp.~109--125, 2013.

\bibitem{Song2009Time}
L.~Song, M.~Kolar, and E.~P. Xing, ``Time-varying dynamic bayesian networks.,''
  {\em Advances in Neural Information Processing Systems}, vol.~22,
  pp.~1732--1740, 2009.

\bibitem{Bouchaala2010Improving}
L.~Bouchaala, A.~Masmoudi, F.~Gargouri, and A.~Rebai, ``Improving algorithms
  for structure learning in bayesian networks using a new implicit score,''
  {\em Expert Systems with Applications}, vol.~37, no.~7, pp.~5470--5475, 2010.

\bibitem{Cheng2002Learning}
J.~Cheng, R.~Greiner, J.~Kelly, D.~Bell, and W.~Liu, ``Learning bayesian
  networks from data: An information-theory based approach,'' {\em Artificial
  Intelligence}, vol.~137, no.~12, p.~43¨C90, 2002.

\bibitem{Wang2011Time}
Z.~Wang, E.~E. Kuruoglu, X.~Yang, and Y.~Xu, ``Time varying dynamic bayesian
  network for nonstationary events modeling and online inference,'' {\em IEEE
  Transactions on Signal Processing}, vol.~59, no.~4, pp.~1553--1568, 2011.

\bibitem{Murphy2002Dynamic}
K.~P. Murphy, ``Dynamic bayesian networks: Representation, inference and
  learning,'' {\em Probabilistic Graphical Models}, vol.~13, pp.~303 -- 306,
  2002.

\bibitem{Robinson2008Non}
J.~W. Robinson and A.~J. Hartemink, ``Non-stationary dynamic bayesian
  networks.,'' {\em Advances in Neural Information Processing Systems},
  vol.~11, no.~18, pp.~1369--1376, 2008.

\bibitem{2012Data}
``Hardware hacking for data scientists.'' \url{http://datasensinglab.com//}.
\newblock 2012.

\end{thebibliography}
\end{document}